\documentclass[twocolumn,english,aps,pra]{revtex4-2}

\usepackage{array}
\usepackage[latin9]{inputenc}
\usepackage{amsmath}
\usepackage{amssymb}
\usepackage{graphicx}
\usepackage{babel}
\usepackage{mathrsfs}
\usepackage{amsfonts}
\usepackage{epstopdf}
\usepackage{makecell}
\usepackage{multirow}
\usepackage{color}
\usepackage{natbib}
\usepackage{bm}
\usepackage{textcase}

\begin{document}
\title{Topological Quantum Criticality in Superfluids and Superconductors:\\
Surface criticality, Thermal properties, and Lifshitz Majorana fields }

\author{Fan Yang and Fei Zhou}

\affiliation{Department of Physics and Astronomy, University of British Columbia, Vancouver, British Columbia, V6T 1Z1, Canada}

\date{\today}

\begin{abstract}

Time reversal invariant (TRI) topological superfluids  (TSFs) and topological superconductors (TSCs) are robust symmetry protected gapped topological states.
In this article, we study the evolution of these topological states in the presence of time reversal symmetry breaking (TRB) fields and/or sufficiently large TRI fields.
Physically, one of the realizations of TRB fields can be internal spin exchange fields due to background magnetic ordering.
We find that the fully gapped TSFs and TSCs are generically separated from other nodal states by various zero temperature quantum critical points that are characterized by generalized quantum Lifshitz Majorana fields with distinct scaling properties. These emergent Lifshitz Majorana fields also define finite temperature properties in quantum critical regimes.
Moreover, for a certain subset of TRB fields, there exists a precursor to bulk transitions, 
where surface states can also exhibit quantum critical behavior near zero fields.   
\end{abstract}

\maketitle

\section{Introduction}

Topological superfluids and superconductors have been a fascinating subject for long.
Early explorations of topological superfluids were at least partially motivated by their close connections to quantum Hall physics \cite{Volovik88,Volovik}. 
Read and Green further pointed out the unique roles played by Majorana edge states in time reversal symmetry breaking (TRB) topological states and in phase transitions between topological and non-topological states \cite{Read00}.
Possible non-abelian statistical properties in topological states \cite{Moore91,Nayak96,Read96,Ivanov01} have made topological superfluids and superconductors one of the very promising 
candidates for topological quantum computers, an idea put forward by Kitaev \cite{Kitaev01,Kitaev03}.

Time reversal invariant (TRI) topological superfluids and superconductors studied in more recent literatures are relatively young members of topological states \cite{Roy08,Qi09,Qi11,Bernevig, Zhang13,Mizushima16,Sato17}. 
These studies are also related to the developments in topological insulators \cite{Kane05,Bernevig06a,Bernevig06b,Fu07,Fu07b,Moore07,Qi08, Hasan10}.
The possibility of having topological superconducting states in heterostructures has also generated enormous excitements and interest \cite{Fu08,Qi10a,Lutchyn10,Chung11,Nakosai12}. 
Impressive efforts have been made to systematically classify these states and characterize them in terms of elementary Fermi surface properties \cite{Schnyder08,Kitaev09,Qi10,Teo10}. 
These research efforts generalize the notion of topological states beyond the previously known examples of topological superfluids
or superconductors and perhaps provide very broad searching criteria in potentially realizing them in quantum materials. They open a door to many new studies of topological matter, both theoretical and experimental.

More recently, there has also been growing interest in nodal topological superfluids and superconductors. 
In these cases, topological invariants can be defined on momentum space submanifolds enclosing the nodes \cite{Sato06,Beri10}.
Various nodal structures, such as point nodes, line nodes and surfaces nodes have been investigated and the nodal phases are classified by symmetries of Hamiltonians \cite{Zhao13, Kobayashi14, Zhao16}.
For example, in analogy to Weyl semimetals \cite{Wan11, Burkov11, Burkov18, Armitage18} where topologically protected point nodes exist on Fermi surfaces, Wely superconductors with similar point nodes have been proposed to exist when TRS is broken \cite{Meng12,Cho12,Yang14,Bednik15}. 
Meanwhile, topological properties such as surface states of topological phases with line nodes have been discussed mostly in the context of noncentrosymmetric superconductors \cite{Yada11, Sato11, Schnyder11,Schnyder12}.
Possible realizations of superconducting phase with surface nodes have also been proposed in multiband systems with broken TRS \cite{Agterberg17}.
 
In superfluids and superconductors, there can be phase transitions between various gapped phases, or between gapped and gapless nodal phases that have the same local order parameters and break the same symmetries spontaneously but differ in global topology. 
{``\em How does one phase make a phase transition to  another topologically distinct phase?''} is the key question we want to explore in this study.
These transitions are obviously beyond the standard Landau paradigm of order-disorder phase transitions \cite{Landau} and usually occur when various symmetries such as gauge symmetries are still spontaneously broken.
How are they different from the order-disorder transitions and what are the upper critical dimensions of these transitions, below which strong correlations can emerge?

More concretely, TRI topological superfluids and superconductors are robust, gapped states that are well protected  and
their surface or edge states remain gapless if weak external fields are also time reversal invariant \cite{Mizushima16}. 
However, in the presence of TRB fields, either due to spin exchange fields or pairing exchange effects, surface states can be gapped at any finite TRB fields.
The bulk of a gapped topological state can either simply have a smooth crossover to a gapped topologically trivial superfluid or superconducting state, or in more generic cases that we will focus on below undergo phase transitions to nodal phases with various distinct nodal structures that can also be topological states. 

The main motivation of this article is to identify these quantum transitions that turn out to only exist at $T=0$, i.e. at zero
temperature, and hence can be conveniently characterized as quantum critical points (QCPs) \cite{Sachdev}.  These QCPs also naturally define scaling properties of thermal states in quantum critical regimes.
Phenomenologically, we can always classify the nodal phases into at least three categories: (A) nodal point phases (NPPs), (B) nodal line phases (NLPs), and (C) nodal surface phases (NSPs).
A gapped topological state can undergo a continuous phase transition into one of these phases and correspondingly there should be at least three different universality classes specifying these transitions.

Here, we do not attempt to have an exhaustive classification of all possible QCPs that may exist in superfluids or superconductors.
Rather, we will focus on QCPs that are characterized by three classes of emergent extended quantum Lifshitz Majorana fields (QLMF) with distinct scaling properties. 
Real Majorana fields appear naturally because of breaking of gauge symmetries at the points of transitions; while Lifshitz fields \cite{Lifshitz41} are induced as precursors of nodal structures.  
These QLMFs describe a large variety of QCPs between gapped and nodal phases. We will attribute three types of quantum fields: QLMFA, QLMFB, and QLMFC to transitions from a gapped phase into (A) NPP,  (B) NLP, and (C) NSP, respectively.  
These three fields which all break the Lorentz symmetry, together with relativistic quantum Majorana fields with full emergent Lorentz symmetry appear to form a set of quantum fields naturally emerging at generic QCPs in topological superfluids (TSFs) and topological superconductors (TSCs). 
They capture the low energy physics of a very broad set of QCPs that can exist in TSFs and TSCs.  The emergent Lorentz invariant Majorana fields have been pointed out to appear at 
a beyond-Landau paradigm transition between a TSF and a non-topological superfluid that spontaneously break the same symmetries \cite{Yang19}.
  
Other beyond-Landau paradigm quantum phase transitions have also been studied in various models.
In 2D and 3D topological insulators, topological quantum criticality has been discussed recently \cite{BJYang14, Cho16,Isobe16}.  
One of the most celebrated examples of beyond-Landau paradigm transitions is a class of quantum phase transitions between states with different order parameters or spontaneously breaking different types of symmetries \cite{Senthil04a,Senthil04b}. It was suggested that such QCPs, denoted as {\it deconfined} QCPs, can possess new emergent symmetries and novel particles which do not exist in either of the ordered
phases separated by the QCP. Scale invariant quantum spin liquids can naturally appear at deconfined QCPs, which adds an interesting new member to the earlier family of gapped spin liquids and quantum spin valence bond solids \cite{Haldane88,Rokhsar88,Read91,Moessner01,Senthil04a,Senthil04b}.

The topological quantum phase transitions discussed in this article are driven by changes in global
topology as mentioned before. However, the unique feature that distinguishes them from other beyond-Landau paradigm transitions is that U(1) symmetry is broken spontaneously and charges are not conserved at these topological QCPs in superfluids and superconductors. 
This aspect plays a paramount role in the following construction of effective field theories for these QCPs.

The rest of this article is organized as follows. In Sec. \ref{general}, we introduce the basic notions of Majorana fields and QLMFs, and general phenomenologies of three classes of  QCPs described by these QLMFs. 
We will present the main results, such as the order of phase transitions and the existence of surface QCPs with TRB fields.
In Sec. \ref{pwave}, we will focus on the applications to the simplest $p$-wave superfluids and further zoom in to examine QCPs specifically for $p$-wave superfluids at $T=0$.
In Sec. \ref{TQCP}, we discuss the properties of these QCPs at finite temperature.
In Sec. \ref{Dirac}, we generalize our studies to TSCs with extra orbital degrees of freedom.
We present applications to TSCs of Dirac fermions and classify all possible fields that can drive topological quantum phase transitions.
In Sec. \ref{CBS}, as another application we further present results on QCPs in TSC of Cu$_x$Bi$_2$Se$_3$ that was pointed out previously \cite{Fu10, Sasaki11}. 

\section{QLMF\lowercase{s} and quantum criticality}\label{general}

\subsection{General Phenomenology at QCPs}

As stated in the Introduction, the specific transitions of our interest in the current article, as well as in a previous article by the authors (Ref. \cite{Yang19})
are either entirely driven by the change of global topologies or involve drastic changes of global topologies. They are either
entirely absent in conventional non-topological superconductors or distinctly different from their counterparts in conventional superconductors.

More importantly, all these QCPs {\em do not} appear in the Landau paradigm of order-disorder phase transitions because the states on both sides of the QCPs and the QCPs themselves all break the same symmetries spontaneously. All transitions considered below and in Ref. \cite{Yang19} do not involve condensation of new bosonic quantum fields or particles.
It is exactly the global topology and topological distinctions between different states that result in such a new class of QCPs.
For this reason, we refer them as topological QCPs.

To illustrate the possibility of a continuum quantum field theory representation of topological quantum criticality, we apply an adiabatic theorem to the situation under our consideration. A gapped fermonic topological state (allowed by global symmetries) can maintain its topological distinction under small Hamiltonian deformations if the gap remains open. So a change in global topologies requires closing of a gap.
The gap closing, which is necessary for changes of global topologies, indicates coalescing of gapped particles into the ground state near topological QCPs.

The gap-closing phenomena in topological phase transitions have some unique features compared with those in many conventional non-topological quantum phase transitions. In topological superconductors and fermonic superfluids, elementary emergent particles are fermonic. In fact, they are real Majorana fermions due to the emergent charge conjugation symmetry at U(1) symmetry breaking QCPs. So generically, these topological quantum phase transitions occur with coalescing of real fermions into the ground state without {\em condensation} of new quantum fields or spontaneously breaking additional symmetries. This aspect is obviously fundamentally different from the Landau paradigm of order-disorder transitions.
On the other hand, the above observation does explicitly suggest an effective quantum (Majorana) field theory representation for coalesce dynamics of gapped particles and therefore topological quantum criticality.

Furthermore, different topologies of quantum phases naturally require different quantum field theory representations and hence result in different universality classes. For instance, a nodal point topology demands a very different quantum Lifshitz field theory
than a nodal line topology would demand in the bulk. 
Meanwhile, perhaps the most remarkable consequence of the gap closing in topological states is the proliferation of surface states into the interior of topological matter. This is essential in topological transitions so that a state can topologically reconstruct in the bulk and boundary simultaneously across a transition. Physically, these bulk transitions are always accompanied or even heralded by surface quantum criticality, a very distinct aspect of topological quantum criticality. These two particular aspects, one reflecting the bulk topology and the other more on its consequences at boundaries, are both absent in other gap closing phenomena in conventional non-topological systems as well as in the standard Landau paradigm. Actually these two aspects are what makes the topological QCPs outstanding. Therefore, any appropriate quantum field theory representation for topological QCPs have to be in a class of quantum field
theories with robust boundary states reflecting a corresponding change of topologies across the QCPs. This is indeed one of the guiding principles in our explicit constructions below, apart from more standard symmetry considerations.
The unique role of changes of topologies in these transitions is therefore encoded in these quantum fields.

An alternative and more practical way to think about these unique transitions is to compare them with what happens in a non-topological $s$-wave superconductor when one varies the same parameters.
For instance, if one increases the interaction strength from weak to strong, there are no transitions in conventional $s$-wave superconductors, because the superconductor has the same type of local order or break the same U(1) symmetry. This is not the case in
topological superconductors because there can be a change of global topology
when interactions increase, even though states are the same locally \cite{Yang19}. This results in a bulk transition (although with relatively higher order), in addition to surface quantum transitions. So the very existence of such transitions
entirely and crucially relies on the notion of global topology of underlying superconductors and topological distinction of different states. In other words, global topologies and their characterization add a new dimension to the parameter space along which phase transitions can occur. This new dimension is beyond the standard Landau paradigm.

Because of broken gauge symmetry, a generic transition from a gapped topological phase to nodal phases in TSFs and TSCs can be characterized by an effective QLMF, i.e., dynamics of real fermions.
The emergence of nodal structures in the low energy limit after transitions generally requires different scaling properties along different spatial-temporal directions near QCPs.
This suggests the relevance of extended quantum Lifshitz Majorana fields. 

For transition to NPPs, the effective field theory construction that takes into account the nodal point feature suggests the following universality class that we name QLMFA. In QLMF theories, the fundamental fields are $2N$-component Majorana fields or real fermions defined as
\begin{eqnarray}
&& \chi({\bf x}) =(\chi_1, \chi_2, ...,\chi_{2N})^T, \nonumber \\
&& \chi^\dagger_i ({\bf x}) =\chi_i({\bf x}), \nonumber \\
 && \{\chi_i({\bf x}),\chi_j({\bf x})\}=\delta({\bf x}-{\bf x'}) \delta_{ij}, \quad i, j=1,2,...,2N.
\end{eqnarray}
In terms of fundamental Majorana fields, QLMFA has the following simple generic structure in $d$ dimensions,
\begin{eqnarray}\label{HA}
&&{  H}_\text{A} =\frac12\int d^d{\bf x}\chi^T [ \Gamma_1 (-\partial^2_{x_1} + m)  +\sum_{\alpha=2}^{d} \Gamma_\alpha i \partial_{x_\alpha} ] \chi +{  H}_I,\nonumber\\
&& \{ \Gamma_\nu,\Gamma_\rho \}=\delta_{\nu,\rho},\quad  \nu,\rho=1,2,...,d, \nonumber \\
&& \Gamma_1^\dagger=\Gamma_1=-\Gamma^T_1, \quad \Gamma_\alpha^\dagger=\Gamma_\alpha=\Gamma_\alpha^T,
\end{eqnarray} 
where $\Gamma_\alpha$, $\alpha=2,...,d$, are $2N\times 2N$ real symmetric Hermitian matrices while
$\Gamma_1$ is a purely imaginary antisymmetric Hermitian matrix. $\Gamma_\nu$, $\nu=1,2,...,d$, all anti-commute with each other.
These symmetries of the $\Gamma$ matrices are to preserve the charge conjugation symmetry of Majorana fermions. 
In general if $N'$ is the number of bands involved, we should have $N' \geq N \geq 1$. Detailed structures of $\Gamma$ matrices will be shown in the following sections for specific models.
$m$ is the mass term defining the transition. At QCPs, we have $m=0$. 

${H}_I$ is the interactions between Majorana fermions and background dynamic fluctuations represented by real scalar fields $\phi_\gamma$, $\gamma=1,2,...Q$. 
The coupling can be conveniently expressed as
\begin{eqnarray}
&& {  H}_I ={  H}_\phi + {  H}_{\phi\chi}, \nonumber \\
&& {  H}_\phi=\frac12\int d^d{\bf x}\sum_\gamma [ \pi^2_\gamma + (\nabla \phi_\gamma)^2 +m_\phi^2 \phi^2_\gamma ],  \nonumber \\
&& {  H}_{\phi\chi}=\int d^d{\bf x} \sum_{i,j,\gamma}\phi_\gamma \chi_i g^\gamma_{ij} \chi_j, \nonumber\\
&&  g^\gamma_{ij}=-g^\gamma_{ji}, \quad {g^\gamma_{ij}}^*=-g^\gamma_{ij}, \nonumber\\
&& [\phi_\gamma({\bf x}), \phi_\zeta({\bf x')}]=0,\quad [\phi_\gamma({\bf x}), \pi_\zeta({\bf x'})] =i \delta_{\gamma\zeta} \delta({\bf x}-{\bf x'}).\nonumber\\
\end{eqnarray} 
Here $g^\gamma_{ij}$ are elements of $G^\gamma$, a set of purely imaginary anti-symmetric tensors that couple Majorana fermion field $\chi_i$, $i=1,2,...,2N$, to the real field $\phi_\gamma$, $\gamma=1,2,...,Q$. $\pi_\gamma$ is the momentum field conjugate to $\phi_\gamma$.

In the limit of massive scalar fields and $2N \geq 4$, 
${  H}_I$ in the low energy limit can be further cast into an effective form of four-fermion operators,
\begin{eqnarray}
&& {  H}_I= \int d^d{\bf x}\sum_{i,j,k,l} \chi_i \chi_j \chi_k \chi_l \Pi_{i,j,k,l}+..., \nonumber \\
&& i,j,k,l=1,2,...,2N,
\label{4F}
\end{eqnarray}
where $\Pi_{i,j,k,l}$ is an antisymmetric tensor under the interchange of any pair of indices, e.g.
$\Pi_{i,j,k,l}=-\Pi_{j,i,k,l}=-\Pi_{i,k,j,l}=-\Pi_{i,j,l,k}$, etc.
Only four-fermion interaction terms, which are most relevant, are present here; other less relevant terms are muted in the form of the ellipsis.
However, for $2N=2$ the four-fermion (local) operator vanishes and minimal models must involve dynamic fields $\phi$.

In the same limit, the Hamiltonian ${  H}_\text{A}$ has an emergent scale symmetry at a QCP if one introduces the following scale transformation
\begin{eqnarray}
&& t \rightarrow t'=\lambda^2 t, 
\nonumber \\ 
&& x_1 \rightarrow {x'_1}= \lambda x_1, \nonumber \\
&& x_\alpha \rightarrow {x'_\alpha} =\lambda^2 x_\alpha, \quad \alpha=2,..,d,
\end{eqnarray}
and the Majorana field transforms accordingly
\begin{eqnarray}
\chi ({\bf x}) \rightarrow \chi'({\bf x'})=\lambda^{-\eta_A} \chi({\bf x}),
\end{eqnarray}
with $\eta_A=d-1/2$ for a free field theory QCP.

By the same token, we list the main properties of QLMFB which can characterize the QCPs for transitions into NLPs,
\begin{eqnarray}
&& {  H}_\text{B} =\frac12\int d^d{\bf x}\chi^T [ \Gamma_1 (-\partial^2_{x_1}-\partial^2_{x_2} +m )  + \sum_{\alpha=3}^{d} \Gamma_\alpha i \partial_{x_\alpha}] \chi +{  H}_I \nonumber \\
&& \{ \Gamma_\nu,\Gamma_\rho \}=\delta_{\nu\rho},\quad \nu,\rho=1,2,..., d. \nonumber \\
&& \Gamma_{1}^\dagger=\Gamma_{1}=-\Gamma^T_{1},\quad \Gamma_\alpha^\dagger=\Gamma_\alpha=\Gamma_\alpha^T, \quad \alpha=3,...,d.
\end{eqnarray} 
The Hamiltonian is scale invariant at a QCP if one introduces the following scale transformation
\begin{eqnarray}
&& t \rightarrow t'=\lambda^2 t, 
\nonumber \\ 
&& x_{1,2} \rightarrow {x'_{1,2}}= \lambda x_{1,2}, \nonumber \\
&& x_\alpha \rightarrow {x'_\alpha} =\lambda^2 x_\alpha, \quad \alpha=3,...,d,
\end{eqnarray}
and the Majorana field transforms accordingly
\begin{eqnarray}
\chi ({\bf x}) \rightarrow \chi'({\bf x'})=\lambda^{-\eta_B} \chi({\bf x}),
\end{eqnarray}
with $\eta_B=d-1$ for a free field theory QCP.

Finally, one can also have QLMFC which only involves one antisymmetric $\Gamma$ matrix,
\begin{eqnarray}
&& {  H}_\text{C} =\frac12\int d^d{\bf x}\chi^T \Gamma_1 (-\nabla^2 +m) \chi +{  H}_I \nonumber \\
&& \Gamma_{1}^\dagger=\Gamma_{1}=-\Gamma^T_{1}.
\end{eqnarray} 
The scaling property of QLMFC is identical to a non-relativistic field theory with Galilean invariance. However it has an additional charge conjugation symmetry of Majorana fermion
and therefore always couple to an antisymmetric purely imaginary $\Gamma$ matrix.
The Hamiltonian is scale invariant at a QCP if one introduces the following scale transformation
\begin{eqnarray}
&& t \rightarrow t'=\lambda^2 t, 
\nonumber \\ 
&& x_{\alpha} \rightarrow {x'_{\alpha}}= \lambda x_{\alpha}, \quad \alpha=1,..,d,
\end{eqnarray}
and the Majorana field transforms accordingly
\begin{eqnarray}
\chi ({\bf x}) \rightarrow \chi'({\bf x'})=\lambda^{-\eta_c} \chi({\bf x}),
\end{eqnarray}
with $\eta_C=d/2$ for a free theory QCP.

At last, the four-fermion interaction $\Pi$ in Eq. (\ref{4F}) transforms as 
\begin{eqnarray}
\Pi \rightarrow \Pi'=\lambda^{2\eta_{A,B,C}-2} \Pi.
\end{eqnarray}

The above equation indicates that the upper critical dimensions $D_u$ for interaction operator in which ${  H}_I$ becomes a marginal operator is given by $2\eta_{A,B,C}=2$.
So for QLMFA, QLMAB and QLMFC,  $D_u=3/2$, $2$ and $2$ respectively. 

In contrast, Majorana fermions with Lorentz symmetry naturally appear in quantum phase transitions between topological superfluids and non-topological superfluids with upper critical dimension $D_u=1$ \cite{Yang19}.
The QLMF classes with symmetries lower than the Lorentz symmetry generally have higher upper critical dimensions. 
Especially in the case of QLMFB and QLMFC classes, the QCPs in 2D are strong coupling implying conformal fields of Majorana fermions.

Before ending this part of the discussion, let us comment on the subtle role of global symmetries on the construction of our QLMF theories. As all the topological transitions studied here occur in a U(1) symmetry breaking state, QCPs here always have the charge conjugation symmetry $\mathcal{C}$,
which directly suggests a Majorana representation. But parity symmetry $\mathcal{P}$ or time reversal symmetry $\mathcal{T}$ is usually broken at QCPs, as transitions happen only when external exchange fields that break one or both of these two symmetries are applied.
As we are mainly interested in transitions to various nodal phases that belong to symmetry protected topological (SPT) phases \cite{Chen10, Chen12, Chen13}, we have assumed that the additional global
symmetries, beyond $\mathcal{T}$ or $\mathcal{P}$ symmetries, can also be present in specific quantum matters to physically protect those topological phases (see Sec. \ref{Dirac}). The effective field theories of quantum Lifshitz Majorana fermions are introduced in this particular context if the phases are protected by appropriate global symmetries and corresponding phase transitions do exist. 
So the number of components of Majorana fermions involved or the dimension of $\Gamma$ matrices further depend on the concrete global symmetries needed to define specific stable nodal phases.
However, if the only global symmetry at transitions, apart from the basic charge conjugation symmetry, is $\mathcal{P}$ or $\mathcal{T}$ and a nodal phase is indeed well protected by one of these two symmetries, then QCPs shall only be described by an effective QLMF model with a definitive $N$. 

For instance, if a nodal phase is fully protected by $\mathcal{P}$ symmetry with $\mathcal{T}$ symmetry broken such as in some NPPs, QCPs should be expected to possess $\mathcal{P}$ symmetry only. The transition to such an NPP, if occurs, should be described by QLMF models with $N=1$
only. The QCPs of $N=1$ QLMFs represent stable gapless states as long as $\mathcal{P}$ symmetry is present. In other words, QCPs of $N=1$ are protected by the global symmetry $\mathcal{P}$.

In the same context where only $\mathcal P$ symmetry is present, quantum critical states implied by QLMF models with $N\geq2$ do exist but they are expected to be unstable and their existence requires further fine tuning of multiple relevant terms. In the presence of those relevant $\mathcal P$ symmetric perturbations, we anticipate that QCPs with $N\geq2$ collapse to the universality of $N=1$ QLMFs and this defines the universality classes of transitions with both $\mathcal C$ and $\mathcal P$ symmetries but with $\mathcal T$ symmetry broken. This aspect is equivalent to a general consensus that the universality only depends on symmetries
and is independent of representations of a symmetry group. If the only global symmetries are $\mathcal C$ and $\mathcal P$, without other protecting symmetries QLMF models with $N\ge2$ shall be better considered as appropriate models for topological {\it multicritical} points rather than conventional QCPs.

The other possibility is that gapless states in $N\geq2$ QLMF models are fully gapped in the presence of relevant perturbations; but this perhaps further implies the corresponding nodal phases no longer exist and there are no transitions at all.
In the rest of the discussion, however, we will always assume the gapless nodal phases involved are sufficiently protected by additional global symmetries and so the transitions have to occur. 
The same global symmetries should also be naturally present at QCPs to be consistent. This is encoded implicitly in the dimension $2N\geq4$, as well as the structure of anticommunting $\Gamma$ matrices.
From this point of view, we will simply treat QLMFs with general $N$, $N\geq2$ as different QCPs in the presence of different topologically protecting symmetries in quantum matters. We will come back to this issue when discussing concrete models.

\subsection{General Scaling Properties}

The emergent QLMFs imply unique scaling properties at the QCPs and determine the order of phase transitions.
For ${  H}_\text{A,B,C}$ defined up to a ultra-violet (UV) {\it energy} scale $\Lambda$, the mass and temperature dependence of the grand potential $\Omega$ can always be expressed as
\begin{eqnarray}
&& \frac{\Omega}{V} =\Lambda^{\eta_0+1} \tilde{\Omega}( \tilde{m}, \tilde{{  G}}; \tilde{T}),
\end{eqnarray}
where $V$ is the volume of the system, $\eta_0$ takes the values of $\eta_{A,B,C}$ for free fields of QLMFA, QLMFB, and QLMFC, respectively. 
$\tilde{m}=m /\Lambda$, $\tilde{T}=T/\Lambda$, $\tilde{{  G}}={  G} \Lambda^{\eta_0-1}$ are the dimensionless mass,  temperature, and interaction tensor, respectively.

As the critical physics given by the infrared behavior of the grand potential should not depend on the UV scale of the effective theory, we can set either $|\tilde{m}|=1$ or $\tilde{T}=1$ leading to scaling properties.
We further associate a fixed point under the scale transformation to a QCP by setting $\tilde{{  G}}={  G}^*$, independent of the UV scale $\Lambda$.
Below the upper critical dimension of QLMFs, a QCP is a strong coupling fixed point with $G^*$ only dependent on the spatial dimensionalities.
While above the upper critical dimensions, the QCPs are described by free theories with ${  G}^*=0$. 
In either case, the above scaling argument indicates that at $T=0$ and near a QCP, the leading non-analytical term of $\Omega$ is given by
\begin{eqnarray}
 \frac{\Omega^\text{NA}}{V} =|m|^{\eta_0+1} \tilde{\Omega}( \text{sgn}(m), {  G}^*; 0).
\end{eqnarray}
Note that  $\tilde{\Omega}$ is a constant of order of unity but may further carry logarithmic dependence of $|m|$ at upper critical dimension. 
The scaling exponents $\eta_0+1$ are universal and independent of details of topological states involved. 

For instance, for QCPs associated with the free theory of QLMFA, $\eta_0+1=d+1/2$; for 3D TSCs and TSFs, the transition can be named as $7/2$th order.
For QLMFB, $\eta_0+1=d$ and in 3D these QCPs are of the third order. 
For QLMFC, $\eta_0+1=d/2+1$ and in 3D QCPs coincide with the well known $5/2$th order 
Lifshitz transitions \cite{Lifshitz60}. 

Although the above analyses on thermodynamics are applicable to QCPs characterized by either free Majorana fermions or strongly 
interacting conformal fields of Majorana fermions, the dynamics such as transport phenomena and hydrodynamics strongly depend on whether the QCPs are strong coupling fixed points or not. 
For this reason, the upper critical dimensions computed in the previous subsection are important. 
We have shown that the upper critical dimensions for different QLMFs are
\begin{eqnarray}
&&D_u=\left\{
\begin{array}{cc}
3/2, & \mbox{QLMFA};  \\
2, & \mbox{QLMFB};\\
2, & \mbox{QLMFC}.
\end{array}
\right.
\end{eqnarray}
Below or at the upper critical dimensions, one should expect an emergence of strong coupling Majorana fixed points, which are an analogue of Wilson-Fisher fixed point in the more conventional order-disorder phase transitions in the standard Landau paradigm \cite{Peshkin}.

At finite temperature and in the quantum critical regime where $T \gg |m|$, thermal and other properties are also dominated by these QCPs.
For instance,
\begin{eqnarray}
 \frac{\Omega}{V} =T^{\eta_0+1} \tilde{\Omega}(0, {  G}^*; 1) +...
\end{eqnarray}
where, as one can easily see, the same set of indices appear in the thermal properties near QCPs.
$\Omega$ discussed above is a non-analytical function of $m$ only at $T=0$ as a result of QCPs. 
For QLMFB and QLMFC in $d=2$, the interactions are marginally relevant and further carry logarithmic singularity of $m$, which indicates (2+1)D conformal field theory fixed points.
However, at any finite temperature, $\Omega$ turns out to be a smooth function of $m$ signifying quantum criticality.

Finally, let us also contrast the discussions above with transitions between topological and non-topological superfluids. Those transitions are described by relativistic Majorana fields with
an emergent Lorentz symmetry. The transition there is always of  $(d+1)$th order and the corresponding index $\eta_0+1=d+1$ differs from all the QCPs discussed here \cite{Yang19}.

In the following sections, we will illustrate these emergent QLMFs at a variety of QCPs in TSFs and TSCs where nodal phases appear.

\subsection{Surface Quantum Criticality}

3D TSFs and TSCs support gapless helical states on its surfaces. Opposite surfaces carry states with opposite handedness that are well separated by the bulk and are effectively decoupled.  
Topological surface states can respond to TRB fields in dramatic ways by opening up a gap at any finite field. When this happens, the surface states can be thought as {\em quantum critical} with respect to these TRB fields. Since the topological states are robust against TRI fields, such surface quantum criticality is unique to certain TRB fields, although not all TRB fields result in gapped surface states immediately.
For those TRB fields which do lead to surface quantum criticality near zero field and well before a possible bulk phase transition, surface quantum criticality can also be thought as a precursor to the later bulk transition.

The effective Hamiltonian for surface criticality can be cast into a simple form of 
\begin{eqnarray}\label{SQCP}
&&{  H}_\text{surf}=\frac12\int d^d{\bf x}\chi^T (s_x i\partial_z -s_z  i\partial_x+m s_y ) \chi +{  H}_\text{SI}, \nonumber\\
&&\begin{split}
{  H}_\text{SI}=&\frac12\int d^d{\bf x}(\pi^2 +\nabla \phi \cdot \nabla \phi  +m_\phi^2 \phi^2)\\
& +g_Y\int d^d{\bf x}\phi \chi^T s_y \chi,
\end{split}
\end{eqnarray}
where $s_\alpha$'s are Pauli matrices in spin space, $\chi^T=(\chi_1,\chi_2)$ is a two-component Majorana field and ${  H}_\text{SI}$ describes the interactions between $\chi$ field and a real scalar field $\phi=\phi^\dagger$ in a Yukawa form.
Note that because $\chi$ only has two components, the four-fermion operator vanishes in the case of surface Majorana fermions and $\chi$ can only interact with dynamic field $\phi$.

On the 2D surface when $m_\phi$ is finite, there is only a free field theory fixed point with $g_Y=0$. The surface criticality at $T=0$ is given by the following cusp structure  in the grand potential similar to the discussions on QCPs in 2D,

\begin{eqnarray}
\frac{\Omega_\text{surf}^\text{NA}}{S}=|m|^3 \tilde{\Omega}_s(\text{sgn}(m),0;0)
\label{SQC}
\end{eqnarray}
where $S$ is the total surface area, $\tilde{\Omega}_s(\tilde{m}, \tilde{g}_Y; \tilde{T})$ is a general function of dimensionless mass $\tilde m$, interaction constant $\tilde g_Y$, and $\tilde T$.
For surface criticality, we further have
\begin{equation}
\tilde{\Omega}_s(\text{sgn}(m),0; 0)=\tilde{\Omega}_s(1, 0; 0),
\end{equation}
since the energy spectrum is symmetric for $\pm m$.

Remarkably, following Eq. (\ref{SQC}) where $m$ represents Zeeman fields, the non-analytical part of surface spin susceptibility is given by
\begin{eqnarray}
\chi_M^\text{NA} =-6|m| \tilde{\Omega}(1,0;0),
\end{eqnarray}
which is even in $m$ but varies linearly as a function of $|m|$.
This is in contrast to the more conventional cases where $\chi_M$ can usually be expanded as an even analytical function of $m$ as $\chi_M\sim\chi_M^{(0)}+\chi_M^{(1)}m^2+...$.

At any finite $T$, the susceptibility is analytical and scales as
\begin{eqnarray}
\chi_M \sim T, 
\end{eqnarray}
as $m$ approaches zero.

In the limit $m_\phi\to0$, there can be further emergent supersymmetries in addition to scale symmetries. We do not pursue this topic in the current article. For the examples of emergent supersymmetries in condensed matter systems, we refer readers to Refs \cite{Lee07,Grover14,Yue10,Zerf16,Li17,Jian17}.

In the following sections, we will illustrate the realization of these QCPs and universality classes in a few different TSFs 
and TSCs. 
In the concrete models we examine below, we intend to demonstrate explicitly the physical parameters that one can vary to drive quantum transitions leading to the emergence of QCPs described in the general phenomenology. 
We will also further connect the effective QLMFs to microscopic Majorana quasi-particles in topological states and further explore the physical consequences at both $T=0$ and at finite temperature quantum critical regimes.

Specifically, we will identify:
(1) a few relevant fields that can lead to potential observation of QLMF physics in TSFs/TSCs;
(2)   detailed structures of anti-commuting $\Gamma $ matrices, which effectively define QLMF in concrete topological states, and the corresponding projection operators that lead to the effective field theory description near QCPs.

In particular, we discuss three physical models where QLMFs can potentially emerge:
(1) topological $p$-wave superfluids in both 3D and 2D in Sec. \ref{pwave}.
(2) topological odd-parity pairing states in Dirac semimetals in Sec. \ref{Dirac}.
(3) Fu-Berg model for Cu$_x$Bi$_2$Se$_3$ in Sec. \ref{CBS}.

\section{Topological QCP\lowercase{s} in $p$-wave superfluid model at zero temperature}\label{pwave}

\subsection{The model}\label{pwavemodel}
The simplest one-band model that supports fully gapped topological states is perhaps the 2D $p+ip$ superfluids of spinless fermions. However, without breaking charge conjugation symmetry the only phase transition in this model happens between two fully gapped phases. The phase transition is driven by chemical potential or interactions, and the effective Hamiltonian near the critical point is Lorentz invariant \cite{Yang19}. 

A minimal model that hosts QLMF QCPs is topological $p$-wave superfluids involving two bands labeled by spin indices.
Let us define Majorana operators
\begin{eqnarray}
&& \chi_{+, s} ({\bf r}) =\frac{1}{\sqrt{2}}\left(c_ s({\bf r})+c^\dagger_ s({\bf r})\right), \nonumber \\
&& \chi_{-, s} ({\bf r})=\frac{1}{i\sqrt{2}}\left(c_ s({\bf r})-c^\dagger_ s({\bf r})\right),
\end{eqnarray}
where $ s=\uparrow,\downarrow$ is the spin index, $c_ s$ and $c_ s^\dagger$ are conventional complex fermionic operators. 
In analogy to Nambu spinors, we define 
\begin{eqnarray}
\chi_{\bf k}=(\chi_{+,\uparrow}({\bf k}), \chi_{+,\downarrow}({\bf k}), \chi_{-,\uparrow}({\bf k}), \chi_{-,\downarrow}({\bf k}))^T.
\end{eqnarray}
Notice in the momentum space, the anticommutation relation of Majorana fermions becomes
\begin{align}
\{\chi_i({\bf k}),\chi_j({\bf k'})\}=\delta({\bf k}+{\bf k'})\delta_{ij}, &&\text{with} && \chi_{\bf k}^\dagger=\chi_{-\bf k}^T.
\end{align}

We start with a TRI $p$-wave superfluid with order parameter $\Delta_p({\bf k})=v{\bf s}\cdot{\bf k}(i s_y)$. Here we fix the gauge such that $v>0$. This choice of order parameter corresponds to a superfluid phase with isotropic gap.
In Majorana representation, the Hamiltonian can be written as
\begin{equation}\label{Hamiltonian}
H=\frac12\sum_{\bf k}\chi_{-\bf k}^T\mathcal{H}({\bf k})\chi_{\bf k}+{H}_I,
\end{equation}
where 
\begin{equation}\label{Hp}
\mathcal{H}({\bf k})=v(-\tau_z\otimes  s_z k_x+\tau_x\otimes I k_y+\tau_z\otimes  s_xk_z)-\tau_y\otimes I(\epsilon_k-\mu).
\end{equation}
Here $\epsilon_k=k^2/2$, $\mu$ is the chemical potential,  $\tau_\alpha$'s are Pauli matrices in the $\chi_+$, $\chi_-$ Majorana space, $s_\alpha$'s are Pauli matrices in spin space, $I$ is the $2\times2$ identity matrix.
As a result of charge conjugation symmetry, all terms of odd powers of ${\bf k}$ are coupled to real matrices and all terms of even odd of ${\bf k}$ are coupled to purely imaginary matrices in $\mathcal{H}({\bf k})$.
All interactions are included in ${H}_I$, and they are irrelevant operators for the transitions and muted for most discussions in 3D.

It is well-known that topological phase transitions happen at $\mu=0$ between a fully gapped topological phase with $\mu>0$ and a non-topological phases  with $\mu<0$ \cite{Qi09}.
The topological phase is protected by TRS with robust gapless helical states on the surface. 
In our previous work (see Ref. \cite{Yang19}), we have identified that these transitions are described by a Lorentz invariant free Majorana field theory near the critical point. 

If the symmetry allows other mass fields such as $s$-wave pairing or spin exchange, it is possible to have other phases with different topology.
Let us classify all possible mass fields by  charge conjugation symmetry $\mathcal C$, time-reversal symmetry $\mathcal T$, and parity symmetry $\mathcal P$. Under these symmetries, the Hamiltonian transforms as
\begin{equation}
\mathcal T\mathcal H({\bf k})\mathcal T^{-1}=\mathcal H(-{\bf k}),\qquad \mathcal{T}^2=-1,
\end{equation}
\begin{equation}
\mathcal C\mathcal H({\bf k})\mathcal C^{-1}=-\mathcal H({-\bf k}),\qquad \mathcal{C}^2=1,
\end{equation}
\begin{equation}
\mathcal P\mathcal H({\bf k})\mathcal P^{-1}=\mathcal H(-{\bf k}),\qquad \mathcal P^2=1.
\end{equation}
In the Majorana representation, we have $\mathcal T=\tau_z\otimes(i s_y) K$ and $\mathcal C=K$, with $K$ being complex conjugate operator.
For the $p$-wave superfluid model, we have $\mathcal P=\tau_y$.

For Majorana fermions, all mass fields must be purely imaginary and antisymmetric to preserve charge conjugation symmetry. In total, there are six possible mass fields. Among them $\tau_y\otimes I$ has been associated with chemical potential.
This leaves us with five other fields that can be attributed to two different types of physics: (1) $s$-wave pairing $\tau_z\otimes s_y$ and $\tau_x\otimes s_y$; (2) Zeeman field term $-{\bf S}\cdot{\bf B}$ which can be written in Cartesian components: $\tau_y\otimes s_x B_x$, $-I\otimes s_y B_y$, and $\tau_y\otimes s_z B_z$, with ${\bf S}$ defined as
\begin{equation}
{\bf S}=(-\tau_y\otimes s_x, I\otimes s_y, -\tau_y\otimes s_z).
\end{equation}
The additional $s$-wave pairings will lead to transitions or crossovers to fully gapped non-topological superconducting states. These transitions or crossovers are not described by QMLFs. For this reason, we only include them in Appendix \ref{sp}.
In contrast, Zeeman fields will lead to NPPs and QCPs belonging to QLMFA universality class. These NPPs are protected by parity symmetry $\mathcal P$. In the following, we will discuss the effect of Zeeman fields in details.

\subsection{Phase diagram and phase transitions due to Zeeman field at $T=0$}\label{EFT}

Being charge neutral, Majorana fields do not directly couple to gauge potential of magnetic fields like electrons do.
But they can couple to Zeeman fields through spin. These Zeeman fields can be either external fields or due to internal spin exchange.
The presence of Zeeman fields in superfluids has two major effects.
First, TRS is broken. Therefore, we would expect changes in topology even when the field is weak. A direct consequence of this is the response of surface modes, which will be discussed in Sec. \ref{surface}.
Second, Zeeman field defines a preferred direction for spins, which breaks the isotropy of the spectrum. This anisotropy eventually leads to NPPs when the Zeeman field is strong enough. 

We first obtain the phase diagram by examining the bulk spectrum in the presence of Zeeman fields. For a Zeeman field ${\bf B}$ along an arbitrary direction, the Zeeman Hamiltonian is
\begin{equation}
H_\text{ZM}=-\frac12\sum_{\bf k}\chi^T_{-{\bf k}}({\bf S}\cdot{\bf B})\chi_{\bf k},
\end{equation}
which leads to an anisotropic spectrum
\begin{equation}
E_{\bf k}^{(\pm)}=\sqrt{v^2{\bf k}^2_\perp+(\sqrt{v^2k_\parallel^2+(\epsilon_k-\mu)^2}\pm B)^2}.
\end{equation}
Here, $k_\parallel$ (${\bf k}_\perp$) is the momentum parallel (perpendicular) to ${\bf B}$, and $B=|{\bf B}|$.
The Zeeman field lifts the spin degeneracy, resulting in two non-degenerate energy bands labeled by the superscript.
The bulk gap closes at point nodes in the lower bands $\pm E_{\bf k}^{(-)}$ at ${\bf k_\perp}=0$ and $v^2k_\parallel^2+(k^2_\parallel/2-\mu)^2=B^2$.

\begin{figure*}
\includegraphics[width=\textwidth]{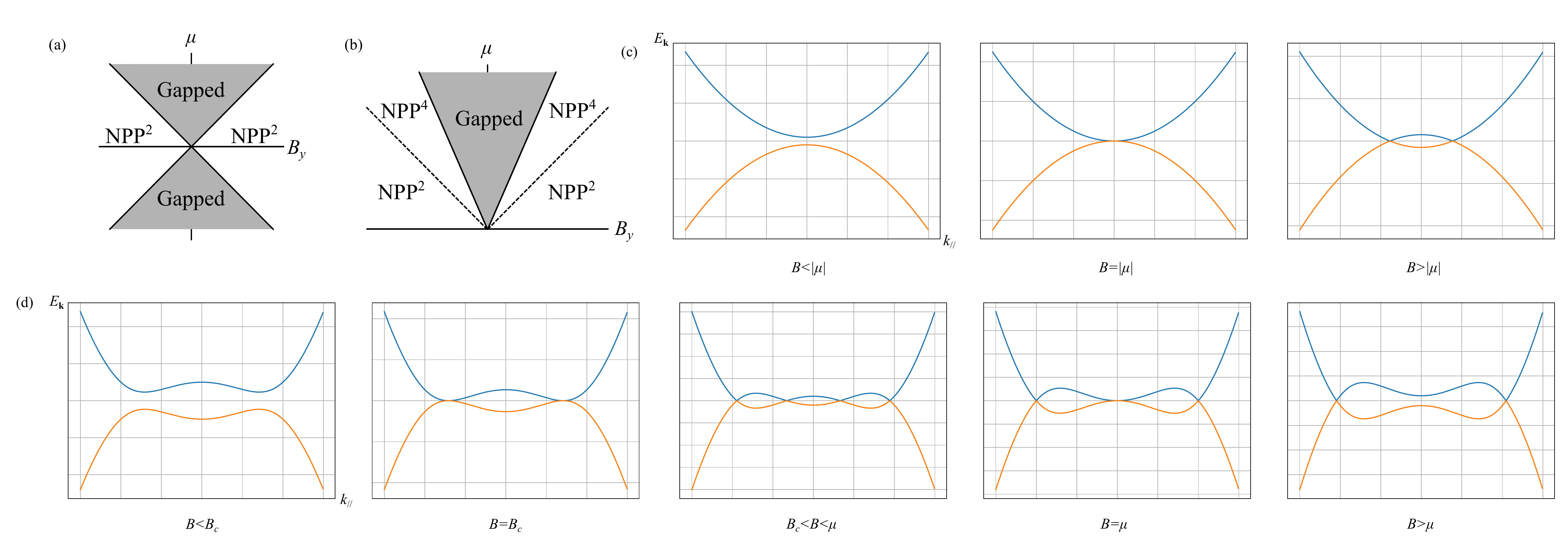}
\caption{Phase diagrams and bulk spectra of 3D $p$-wave superfluids in Zeeman fields.
(a) Phase diagram on the strong coupling side $v^2>\mu$. Without loss of generality, we choose Zeeman field to be along $y$-direction ${\bf B}=B_y{\bf \hat y}$ and $B=|{\bf B}|$. Superfluids are gapped in the shaded area $B<|\mu|$ and gapless with two point nodes in the unshaded area. 
In the nodal point phase (NPP), the number of point nodes are indicated by the superscript.
(b) Phase diagram on the weak coupling side $v^2<\mu$, $\mu>0$.  Superfluids are gapped in the shaded area $B< B_c$ and gapless with point nodes in unshaded area. The dashed lines at $\mu=|B|$ represent the transitions between two different NPPs: NPP with four point nodes (NPP$^4$) for $B_c<B<\mu$ and NPP with two point nodes (NPP$^2$) for $B>\mu$.
(c) Bulk spectrum on the strong coupling side $v^2>\mu$. 
 (d) Bulk spectrum on the weak coupling side $v^2<\mu$. We set ${\bf k}_\perp=0$ and only plot the lower energy bands $\pm E_{\bf k}^{(-)}$ in both (c) and (d).  The higher energy bands $\pm E_{\bf k}^{(+)}$ are always gapped (see text).  \label{phase}}
\end{figure*}

The number of point nodes in the spectrum depends on the ratio $v^2/\mu$  (Fig. \ref{phase}). 
In the $p$-wave superfluid model, $v$ is proportional to the pairing amplitude $\Delta_p({\bf k})$, which further depends on coupling strength between fermions.
Therefore, we need to consider strong ($v^2>\mu$) and weak ($v^2<\mu$) couplings separately.

(1) On the weak coupling side $v^2<\mu$, the chemical potential is always positive. The TRI $p$-wave superfluid is topological in the absence of Zeeman field. 
For given $\mu$, the bulk spectrum suggests two transitions as $B$ increases.
The first one happens at $B_c=\sqrt{\mu^2-(\mu-v^2)^2}$ between a gapped phase and a NPP. At transition $B=B_c$, the bulk gap closes at the two minima of the lower band at ${\bf k_\perp}=0$, $k_\parallel=\pm\sqrt{2(\mu-v^2)}$. 
For weak field $B<B_c$, the bulk is fully gapped. For intermediate field $B_c<B<\mu$, the bulk is gapless with four point nodes at ${\bf k_\perp}=0$, $k_\parallel=\pm\sqrt{2[(\mu-v^2)\pm\sqrt{(v^2-\mu)^2+(B^2-\mu^2)}]}$. 
As the Zeeman field increases further, a second transition happens at $B=\mu$ between two NPPs with different numbers of point nodes. During this transition, the two point nodes in the middle merge into one at ${\bf k}=0$ at the transition and then annihilate each other. The other two point nodes persist.
For strong field $B>\mu$, the bulk has only two point nodes at ${\bf k_\perp}=0$, $k_\parallel=\pm\sqrt{2[(\mu-v^2)+\sqrt{(v^2-\mu)^2+(B^2-\mu^2)}]}$.

The topological quantum phase transitions between gapped phase and NPP in the weak coupling limit are driven by the deformation of Fermi surface and appear to fall outside of the effective field theories listed in Sec. \ref{EFT}. For this reason, we only discuss these transitions in Appendix \ref{weaktransitions}.

(2) On the strong coupling side $v^2>\mu$, $\mu$ can be either positive or negative.  
For given $\mu$, as $B$ increases there is only one phase transition between a gapped phase and a NPP at $B=|\mu|$.  For $B>|\mu|$, the bulk is gapless with two point nodes at ${\bf k}_\perp=0$,  $k_\parallel=\pm\sqrt{2[(\mu-v^2)+\sqrt{(v^2-\mu)^2+(B^2-\mu^2)}]}$. 
The same NPP exists for both positive and negative $\mu$. 
For $B<|\mu|$, the bulk is fully gapped. Note that the TRI topological state with $B=0$ can be smoothly connected to the gapped states with $\mu>B>0$.
In the following, we focus on the phase transitions on the strong coupling side and $\mu>0$.

We illustrate the bulk spectra and phase diagrams for both strong and weak coupling cases in Fig. \ref{phase}.

The NPPs induced by Zeeman field are protected by parity symmmetry $\mathcal P=\tau_y$.

\subsection{Effective field theory for QLMFA transitions}

For the purpose of studying phase transitions, we can use a low energy effective field theory. 
The effective field theory construction relies on the separation between low energy and high energy degrees of freedom. 
This approach focuses entirely on the low energy degrees of freedom after high energy components are integrated out. 
Therefore, the correct effective field theory should include various important renormalization effects due to couplings between high energy and low energy degrees of freedom.

In the following, we construct the effective field theory explicitly for $\mu>0$ in the strong coupling limit and show that it belongs to the QLMFA universality class.

Without loss of generality, we choose Zeeman field to be along $y$-direction. Topological phase transition happens at $B_y=\mu>0$.
In the strong coupling limit $v^2\gg\mu$, we can construct the low energy effective field theory by first projecting the Hamiltonian onto the low energy subspace using projection operator
\begin{equation}
P=P^\tau_{y,+}P^s_{y,+}+P^\tau_{y,-}P^s_{y,-},
\end{equation}
where 
\begin{equation}
P^\tau_{\alpha,\pm}=\frac{I\pm\tau_\alpha}2, \qquad P^ s_{\alpha,\pm}=\frac{I\pm s_\alpha}2.
\end{equation}
Near phase transition, the projected Hamiltonian can be written as 
\begin{equation}\label{strongproj}
\mathcal{H}_\text{proj}({\bf k})=\Gamma_y\Big(\mu-B_y-\epsilon_k\Big)+\Gamma_xvk_x+\Gamma_zvk_z,
\end{equation}
where
\begin{align}
&&\Gamma_x=P(-\tau_z\otimes s_z) P, && \Gamma_z=P(\tau_z\otimes s_x) P,\nonumber\\
&&\Gamma_y=P(I\otimes s_y)P.
\end{align}
The projected Hamiltonian is incomplete for the effective field theory as we also need to include the renormalization effect due to the couplings between high energy and low energy degrees of freedom. After integrating out the high energy degrees of freedom, the leading order effect of these couplings can be written as 
\begin{equation}
\mathcal H^{(2)}({\bf k})=\frac{v^2}{2\mu} k_y^2\Gamma_y
\end{equation}
Combining $\mathcal H_\text{proj}({\bf k})$ and $\mathcal H^{(2)}({\bf k})$ and keeping the leading order term of each momentum component, we find the effective Hamiltonian
\begin{multline}\label{strongeff}
{H}_\text{eff}=\frac12\sum_{\bf k}\chi^T_{\bf -k}\Bigg[\Gamma_y\Big(\mu-B_y+\frac{v^2-\mu}{2\mu}k_y^2\Big)\\
+\Gamma_xvk_x+\Gamma_zvk_z\Bigg]\chi_{\bf k},
\end{multline}
where we have dropped the irrelevant interactions.
This effective Hamiltonian belongs to the universality class of QLMFA. One can recover the phenomenologically constructed Hamiltonian (\ref{HA}) by identifying $m=2\mu(\mu-B_y)/(v^2-\mu)$ and absorbing the coefficients in front of $k_\alpha$ into $\Gamma_\alpha$.

In the strong coupling limit, this low energy effective Hamiltonian is valid for $E_{\bf k}^{(-)}\ll\mu $.

\subsection{Order of phase transitions}
The order of these phase transitions can be computed directly from the zero temperature grand potential 
\begin{equation}
\Omega_0=-\frac12\sum_{{\bf k},i}E_{\bf k}^{(i)},
\end{equation}
where $i$ labels quantum numbers such as spin, orbit, etc.
In the strong coupling limit, the energy spectrum of the effective Hamiltonian is
\begin{equation}\label{effE}
E^\text{eff}_{\bf k}=\sqrt{v^2(k_x^2+k_z^2)+\Big[(\mu- B_y)+\frac{v^2-\mu}{2 \mu}k _y^2\Big]^2}.
\end{equation}
Near phase transitions, the grand potential at $T=0$ is non-analytical with leading non-analytical term
\begin{equation}
\frac{\Omega_0^\text{NA}}V=\frac8{105\pi^2v^2}|B-\mu|^{7/2}\sqrt{\frac{2\mu}{v^2-\mu}}\theta(B-\mu),
\end{equation}
where $\theta(\cdot)$ is the step function.
Therefore, the phase transitions can be named as $7/2$th order, which agrees with the result we obtained using scaling analyses in Sec. \ref{general}.

\subsection{Surface states in Zeeman fields}\label{surface}

To illustrate the change of topology, we study how surface modes respond to Zeeman fields.
We use a cubic geometry such that the superfluids fill in the space $x_0<x<0$, $y_0<y<0$, $z_0<z<0$. It is vacuum in the rest of the space, which can be modeled by setting chemical potential as $\mu\to-\infty$. We focus on the strong coupling side $v^2>\mu$. 

\subsubsection{Weak Zeeman fields}
We first examine the surface states in the gapped phase $B<\mu$, $\mu>0$.
Let us first consider the surface at $y=0$. 
Without Zeeman field, there exist gapless helical surface states,
\begin{gather}
\phi^y_1=\mathcal{N}\begin{pmatrix}1\\ 0\end{pmatrix}_\tau\otimes\begin{pmatrix}\sin({\theta_y}/2)\\\cos(\theta_y/2)\end{pmatrix}_ s e^{\frac1{v}\int_0^y\mu dy}e^{i{\bf k}_\perp\cdot{\bf r}},
\end{gather}
\begin{gather}
\phi^y_2=\mathcal{N}\begin{pmatrix}1\\ 0\end{pmatrix}_\tau\otimes\begin{pmatrix}\cos({\theta_y}/2)\\ -\sin(\theta_y/2)\end{pmatrix}_ s e^{\frac1{v}\int_0^y\mu dy}e^{i{\bf k}_\perp\cdot{\bf r}},
\end{gather}
where $\cot\theta_y=k_x/k_z$ and $\mathcal{N}$ is a normalization factor.
The surface Hamiltonian is gapless 
\begin{equation}\label{surfaceH}
H^{(B=0)}_\text{surf}=\frac12\sum_{\bf k}\psi^T_{-{\bf k},y}(-vk_x s_z+vk_z s_x)\psi_{{\bf k},y},
\end{equation}
where $\psi_{{\bf k},y}=P^\tau_{z,+}\chi_{\bf k}$ is the Majorana operator on this surface. 

In the presence of Zeeman field, it is convenient to define an effective chemical potential at ${\bf k}=0$, 
\begin{equation}\label{mueff}
\mu_\text{eff}^{(\pm)}({\bf k}=0)=\mu\pm B,
\end{equation} 
where the superscript $\pm$ corresponds to the two energy bands $E^{(\pm)}({\bf k})$.
In the fully gapped phase where $0<B<\mu$, $\mu^{(\pm)}_\text{eff}({\bf k}=0)$ is positive for both bands. 
Thus, both bands can support surface states.

However, the surface states can be gapped by the Zeeman field due to broken TRS.
For weak Zeeman field ${\bf B}$ along an arbitrary direction, the surface Hamiltonian for $y=0$ becomes 
\begin{equation}\label{surfaceH}
H_\text{surf}=\frac12\sum_{\bf k}\psi^T_{-{\bf k},y}(-vk_x s_z+vk_z s_x-B_y s_y)\psi_{{\bf k},y},
\end{equation}
up to linear order of $B$. In this linear approximation, the surface modes are gapped by Zeeman field {\it perpendicular} to this surface but not the field parallel to it.
Conversely,  surfaces perpendicular to the Zeeman field become gapped; while the helical Majorana modes on surfaces parallel to the field remain gapless in this approximation (Fig. \ref{3D}). 
This result agrees with the discussions on superfluid ${}^3$He-B in weak magnetic fields in Refs. \cite{Nagato09,Chung09}.

\subsubsection{Strong Zeeman fields}
Next, we examine the surface states in the NPP with strong Zeeman field $B>|\mu|$. Note that in NPP, $\mu$ can be either positive or negative.
Without loss of generality, let us consider Zeeman fields along $y$-direction. 

In the NPP, we have a gapped band $E^{(+)}$ and a nodal band $E^{(-)}$. 
The effective chemical potential $\mu_\text{eff}^{(+)}({\bf k}=0)$ for the gapped band is still positive.
Therefore, the gapped band can still support surface states.
For the nodal band, we can define a generalized momentum-dependent effective chemical potential $\mu_\text{eff}^{(-)}(k_y)$. This effective chemical potential changes sign in momentum space.
Let us label the point nodes as $(0,\pm k_0,0)$, $k_0>0$. The effective chemical potential $\mu_\text{eff}^{(-)}(k_y)$ is positive (negative) for $|k_y|>k_0$ ($|k_y|<k_0$).

The sign of $\mu_\text{eff}^{(-)}(k_y)$ can be argued using the effective Hamiltonian (\ref{strongeff}).
For any given $k_y$, the effective Hamiltonian (\ref{strongeff}) describes a 2D chiral superfluid in the $xz$-plane with Hamiltonian $\mathcal{H}^\text{eff}_{k_y}(k_x,k_z)=\mathcal{H}_\text{eff}({\bf k})$ and effective chemical potential $\mu_\text{eff}^{(-)}(k_y)$. Near transition and near point nodes, we have
\begin{equation}
\mu_\text{eff}^{(-)}(k_y)\approx\mu-B_y+\frac{v^2}{2\mu}k_y^2.
\end{equation}
Here, we have taken the strong coupling limit $v^2\gg\mu$.
In the same limit, we also have
\begin{equation}
k_0^2\approx2\mu(B_y-\mu)/v^2.
\end{equation}
Therefore, $\mu_\text{eff}^{(-)}(k_y)$ is positive (negative) if $k_y^2>k_0^2$ ($k_y^2<k_0^2$) near the point nodes.
Away from these nodes, the band gap of the effective 2D Hamiltonian $\mathcal{H}^\text{eff}_{k_y}(k_x,k_z)$ is given by $\mu_\text{eff}^{(-)}(k_y)$.
Since the band gap only closes at $k_y^2=k_0^2$, the sign of $\mu_\text{eff}^{(-)}(k_y)$ does not change except at $k_y=\pm k_0$.
The sign change at $k_y=\pm k_0$ suggests domain wall structures of $\mu_\text{eff}^{(-)}(k_y)$ in momentum space.

It is well-known that domain wall structures of $\mu$ in real space signal a change of topology across the surface. Similarly, the domain wall structures of $\mu_\text{eff}^{(-)}(k_y)$ in the momentum space signal a change of topology in the momentum space across a surface perpendicular to $k_y$ and containing the point nodes.
If one only focuses on the nodal band $E^{(-)}$, the change of topology in the momentum space would give rise to Fermi arcs for $k_y^2>k_0^2$.
However, to obtain the complete surface states, one needs to also take into account the gapped band $E^{(+)}$.

We can solve the gapless surface states explicitly.
Let us take the surface at $z=0$ as an example.
For $|k_y|<k_0$, only the gapped band can support gapless surface modes 
\begin{equation}\label{phiz}
\phi^z_+=\mathcal{N}(0,1,-1,0)^Te^{\frac1{v}\int_0^z(\mu+B)dz}e^{ik_xx}.
\end{equation}
The surface Hamiltonian is gapless
\begin{equation}
{H}^z_\text{surf}=\frac12\sum_{\bf k}\psi^T_{-{\bf k},z}(vk_x)\psi_{{\bf k},z},
\end{equation}
where the surface Majorana operator is given by $\psi_{{\bf k},z}=(\chi_{+,\downarrow}({\bf k})-\chi_{-,\uparrow}({\bf k}))/\sqrt2$.

For $|k_y|>k_0$, both bands can support surface modes, and we need to take into account the hybridization of these modes.
First, let us consider the zero energy surface modes associated with each band without hybridization. 
The surface states associated with the gapped band is still given by Eq. (\ref{phiz}).
The surface states associated with the nodal band is
\begin{equation}
\phi^z_-=\mathcal{N}(1,0,0,-1)^Te^{\frac1{v}\int_0^z\mu_\text{eff}^{(-)}(k_y) dz}e^{-ik_xx}, \quad |k_y|>k_0.
\end{equation}
Notice that $\phi^z_+$ and $\phi^z_-$ have finite coupling for any $|k_y|>k_0$,
\begin{equation}
\langle\phi^z_+|\mathcal{H}({\bf k})|\phi^z_-\rangle\sim -vk_y+O(k_y^2).
\end{equation}
This coupling results in the hybridization of these two states, and the resultant surface states have finite energy for $|k_y|>k_0$.

\begin{figure}
\includegraphics[width=\columnwidth]{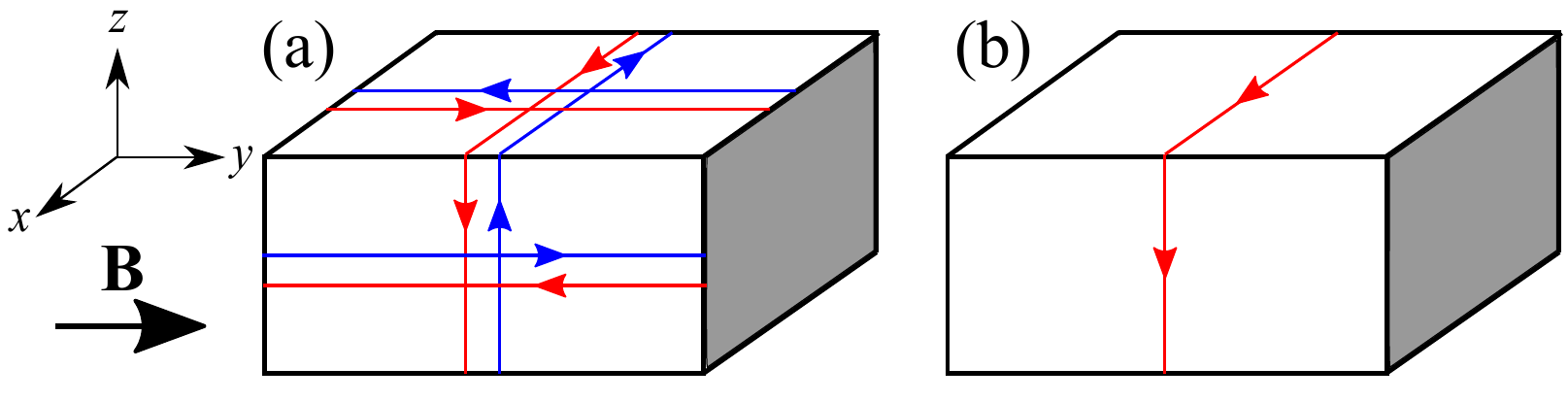}
\caption{Surface states of 3D $p$-wave superfluids in Zeeman field ${\bf B}$. Without loss of generality, we choose ${\bf B}$ to be along $y$-direction. (a) For $B<\mu$, $\mu>0$. The bulk is fully gapped. Surfaces perpendicular to the Zeeman field (shaded) are gapped. Under an approximation up to linear order of $B$, helical modes on surfaces parallel to the Zeeman field remain gapless. (b) For $B>|\mu|$. The bulk is in the nodal point phase. Surfaces perpendicular to Zeeman field remain gapped. Surfaces parallel to Zeeman field have gapless chiral surface states. The chiral surface modes and the Zeeman field form a right-hand grip. \label{3D}}
\end{figure}

Therefore, gapless surface modes only exist between the two point nodes, i.e., at $|k_y|<k_0$. In the momentum space, the zero energy states form a Fermi arc between the point nodes.
In the real space, the gapless surface states are manifested as chiral Majorana modes on surfaces parallel to the Zeeman field. 
Zeeman field and these gapless chiral surface modes on parallel surfaces form a right-hand grip. 
The surfaces perpendicular to the Zeeman field remain gapped (Fig. \ref{3D}).

\subsection{Topological phase transitions on the surface}
\begin{figure}
\includegraphics[width=\columnwidth]{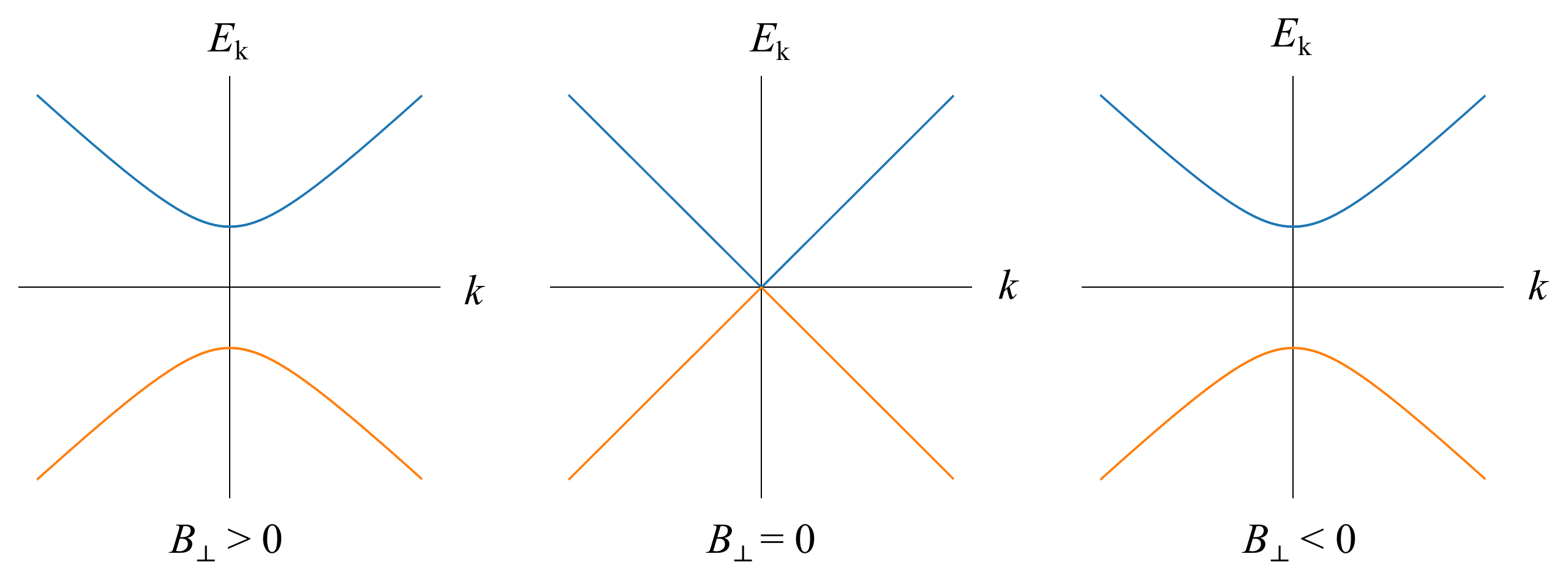}
\caption{Surface spectrum in the presence of perpendicular Zeeman field $B_\perp$. Topological phase transitions happen on the surface between two gapped phases when $B_\perp$ crosses zero. Chern number changes by one across these transitions. \label{stransitions}}
\end{figure}

The surface Hamiltonian is critical at $B_\perp=0$, where $B_\perp$ is the Zeeman field perpendicular to the surface [see e.g. Eq. (\ref{surfaceH})].
On either side of the surface critical point, the surface states are gapped with broken TRS (Fig. \ref{stransitions}). Therefore, one can define a Chern number as the topological invariant for the gapped surface Hamiltonian.
Topological phase transitions between these two gapped phases with different Chern numbers take place on  surfaces as Zeeman field perpendicular to the surface sweeps across zero. 
Across these transitions, the order parameter does not change but Chern number changes by one. These transitions belong to the universality class of Lorentz invariant free Majorana fermions in 2D.

The grand potential for the surface states at $T=0$ is
\begin{equation}
\begin{split}\label{SurfaceOmega}
\frac{\Omega_{0,s}}S&=-\frac12\int\frac{d^2{\bf k}}{(2\pi)^2}\sqrt{v^2k^2+B_\perp^2}\\
&=\frac{|B_\perp|^3}{12\pi v^2}+\text{analytical terms}.
\end{split}
\end{equation}
Thus, the surface topological phase transitions are of the $3$rd order. 

As a result, the surface spin susceptibility has a non-analytical part
\begin{equation}
\chi_M^\text{NA}=-\frac{|B_\perp|}{2\pi v^2},
\end{equation}
which varies linearly in $|B_\perp|$.
The overall susceptibility is an even function of $B_\perp$ with additional analytical terms,
\begin{equation}
\chi_M=\chi_M^{(0)}+\chi_M^\text{NA}+\chi_M^{(1)}B_\perp^2+...,
\end{equation}
where $\chi_M^{(0)}$, $\chi_M^{(1)}$, ... are non-universal and depend on the microscopic properties of the superfluids.

\subsection{$p$-wave superfluids in 2D}
Similar QLMFA QCPs also exist in 2D $p$-wave superfluids. The QCPs are still described by free theory, since the upper critical dimension for QLMFA is $D_u=3/2$.

By setting $k_z=0$ in Eq. (\ref{Hp}), we obtain the Hamiltonian for 2D TRI $p$-wave superfluids in the $xy$-plane. In this case, fermions with $ s_z=+1$ and $ s_z=-1$ are paired in $p_x-ip_y$ and $p_x+ip_y$ channels, respectively.

In 2D, Zeeman fields parallel and perpendicular to the superfluid plane have different effects.

\subsubsection{In-plane Zeeman fields}
Zeeman field parallel to the superfluid plane (in-plane Zeeman field) can drive transitions to NPPs. In the strong  coupling limit, these transitions belong to QLMFA class.
In this case, the effective Hamiltonian is given by Eq (\ref{strongeff}) with $k_z=0$. 
We find, at $T=0$, the leading non-analytical term in the grand potential to be
\begin{equation}
\frac{\Omega^\text{NA}_\text{2D}}S=\frac2{15\pi v}\left|B-\mu\right|^{5/2}\sqrt{\frac{2 \mu}{v^2-\mu}},
\end{equation}
which suggests a $5/2$th order transition.
 
\begin{figure}
\includegraphics[width=\columnwidth]{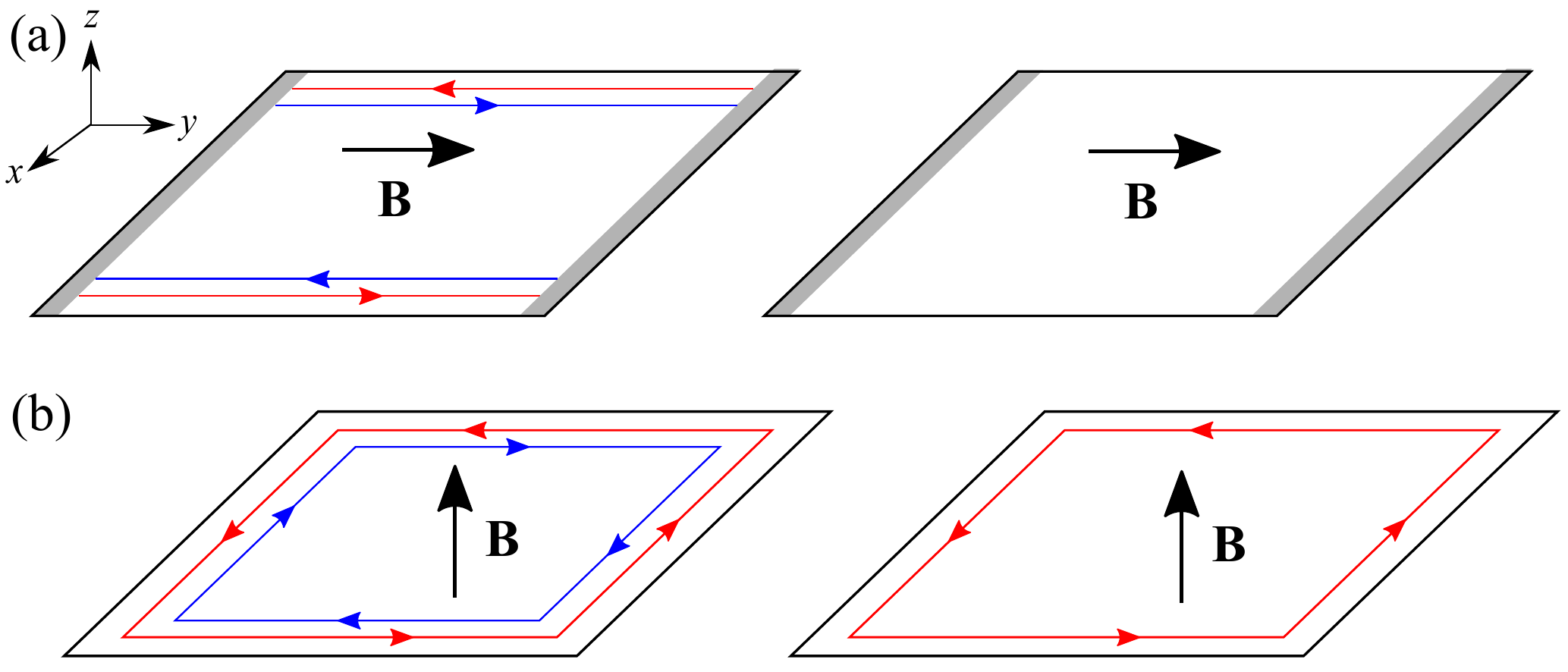}
\caption{Edge states of 2D $p$-wave superfluids in the presence of Zeeman field ${\bf B}$, $B=|{\bf B}|$.  (a) Zeeman field parallel to the superfluid plane (in-plane Zeeman field). For $B<\mu$, $\mu>0$, edges perpendicular to the Zeeman field (shaded) are gapped. Under an approximation up to linear order of $B$, edges parallel to the Zeeman field support counter-propagating gapless edge modes with opposite spins. For $B>|\mu|$, edges perpendicular to the Zeeman field remain gapped. Edges parallel to the Zeeman field support zero-energy flatband  states.
(b) Zeeman field perpendicular to the superfluid plane (out-of-plane Zeeman field).
For $B<\mu$, $\mu>0$, there are gapless helical edge modes on all edges. 
For $B>|\mu|$, there is only one gapless chiral edge mode with spin along the Zeeman field. The chiral edge mode and the Zeeman field form a right-hand grip.
\label{2D}}
\end{figure}

Edge states respond to in-plane Zeeman fields similarly to the 3D case: they are gapped by the Zeeman fields perpendicular to the edge. 
The edge Hamiltonians are critical when the Zeeman field perpendicular to the edge is zero.
Topological phase transitions happen on the edges when the Zeeman field perpendicular to it is tuned across zero. 
Let us take Zeeman field along $y$-direction as an example to illustrate the edge states (Fig. \ref{2D}). For $\mu>0$, $|B_y|<\mu$, the edges perpendicular to the Zeeman field are gapped. 
Under the same linear approximation used for Eq. (\ref{surfaceH}), the edges parallel to Zeeman field have gapless counter-propagating Majarona edge modes with opposite spins.
For $|B_y|>|\mu|$, the edges perpendicular to Zeeman field remain gapped, while each edge parallel to the Zeeman field has a zero-energy flat band.

\subsubsection{Out-of-plane Zeeman fields}
In contrast, Zeeman field perpendicular to the superfluid plane (i.e., along $z$-direction) does not lead to nodal phases. The bulk spectrum
\begin{equation}
E_{{\bf k},z}^{(\pm)}=\sqrt{v^2(k_x^2+k_y^2)+(\epsilon_k-\mu\pm B_z)^2}
\end{equation}
is always gapped except at transitions $B_z=\pm\mu$ when gap closes at ${\bf k}=0$. 

For $\mu>0$ and $|B_z|<\mu$, despite the broken TRS, the helical edge modes are still present. This gapped phase can be smoothly connected to the 2D TRI topological superfluids at $B=0$. 
For $|B_z|>|\mu|$, there exists only one chiral edge mode with spin along $B_z$, which forms a right-hand grip with the Zeeman field. This phase can be smoothly connected to the chiral superfluids in 2D (Fig. \ref{2D}).

\section{Topological quantum criticality}\label{TQCP}
In this section, we will discuss the properties of 3D TSFs/TSCs near the topological QCPs at finite temperature. These discussions not only apply to $p$-wave superfluids discussed in Sec. \ref{pwave}, but also to the TSC models in the following sections.

\subsection{Finite temperature}
First, we show that the transitions in the bulk and on the surface discussed in this article only exist at $T=0$. 
As shown in Sec. \ref{general}, all classes of QLMF QCPs in 3D, as well as the surface QCPs, are described by free field theories. Therefore, we can compute the grand potential near QCPs using simple thermodynamic relations for non-interacting systems. The grand potential $\Omega$ can be split into two parts $\Omega=\Omega_0+F$, where $\Omega_0$ is the zero temperature grand potential and $F$ is the thermal free energy.
Utilizing standard thermodynamic relations for fermions, we have
\begin{equation}
\Omega=-T\sum_{{\bf k},i}\Bigg[\ln\left(1+e^{-\beta E_{\bf k}^{(i)}}\right)+\frac{\beta E_{\bf k}^{(i)}}2\Bigg],
\end{equation}
where $\beta=1/T$.

For the purpose of understanding QCPs, only the infrared behavior of $\Omega$ is relevant.
We can expand the grand potential into a power series of $\beta E_{\bf k}^{(i)}$ as
\begin{equation}\label{OmegaT}
\Omega=T\sum_{{\bf k},i}\sum_{l=1}^\infty\left[\left(\beta E_{\bf k}^{(i)}\right)^{2l}Li_{1-2l}(-1)\right],
\end{equation}
with $Li_l(\cdot)$ being the polylogarithm function.
Notice that this expansion only contains even powers of $E_{\bf k}^{(i)}$. Since $(E_{\bf k}^{(i)})^2$ is an analytical function of $m$ for all the transitions we are interested in, $\Omega$ is also an analytical function of $m$ at any finite temperature. 
Therefore, all transitions discussed in this article, which are described by either QLMFs or Lorentz invariant Majorana fields, only exist at zero temperature. 

\begin{figure}
\includegraphics[width=\columnwidth]{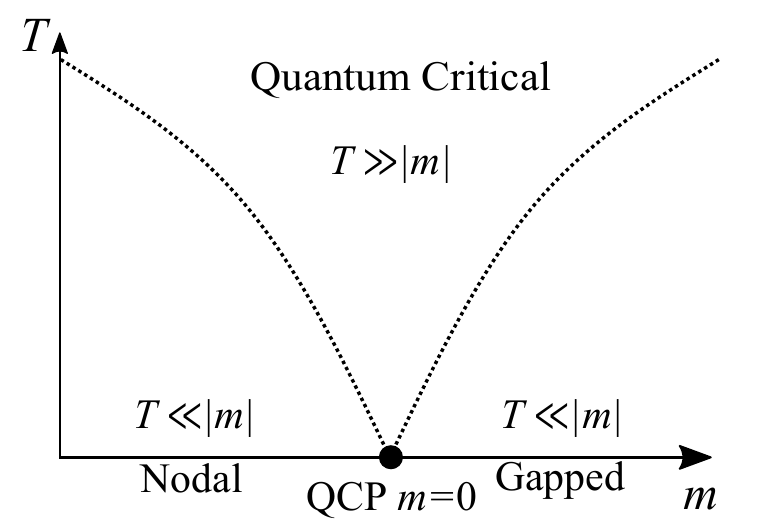}
\caption{Finite temperature phase diagram. Topological phase transitions discussed in this article only exist at zero temperature at $m=0$. $T=0$ and $m=0$ corresponds to a quantum critical point (QCP). In the quantum critical regime where $T\gg |m|$, the temperature scaling of thermodynamic quantities in superfluids and superconductors are universal and dictated by the QCP. Outside the universal quantum critical regime where $T\ll |m|$, the states have similar properties to the zero temperature phases. For QLMF QCPs, the zero temperature phases are gapped (nodal) for $m>0$  ($m<0$). For surface QCPs, we have $m=B_\perp$ and the QCP is at $B_\perp=0$. $B_\perp$ is the Zeeman field perpendicular to the surface. The zero temperature phases on the surface are gapped on both sides of the QCP. \label{QCP}}
\end{figure}

The QCP divides the phase diagram into three smoothly connected regions with qualitatively different behaviors (Fig. \ref{QCP}).
In the quantum critical regime $T\gg|m|$, the temperature scaling of thermodynamic quantities in the superfluids is universal and dictated by the QCP (see also Sec. \ref{general}). The scaling exponents are unique to each universality class. These scaling behaviors can be used to probe the QCPs at zero temperature.
Outside the quantum critical regime where $T\ll|m|$, the low temperature properties of the superfluids are similar to the zero temperature states on either side of the QCP.

\subsection{Surface quantum criticality}\label{SQCPs}
Let us now focus on the surface quantum criticality induced by Zeeman or spin exchange fields.
In this case, $m=B_\perp$ and the QCP is at $B_\perp=0$.
The surface spectrum is fully gapped except at QCP.
We compare the spin susceptibility inside the quantum critical regime $T\gg |B_\perp|$ and outside the regime $T\ll |B_\perp|$.

Inside the quantum critical regime where $T\gg |B_\perp|$, the temperature is much higher than the excitation gap; thermal excitations exhibit universal properties. Thus, susceptibilities should have universal dependence on $T$.
The surface spin susceptibility is analytical in the quantum critical regime and scales linearly in $T$ in the leading order
\begin{equation}
\chi_M\sim T.
\end{equation}

Outside the quantum critical regime where $T\ll |B_\perp|$, the excitation gap is much higher than the temperature. Therefore, all thermal excitations are suppressed exponentially, and the thermal free energy $F$ is exponentially small.
The total grand potential is mainly given by the zero temperature contribution $\Omega\approx\Omega_{0}$.
Therefore, when $|B_\perp|\gg T$, surface spin susceptibility has similarly scaling as the zero temperature case. But different from $T=0$ case, the surface susceptibility is always analytical at finite $T$.

Finally, we would like to emphasize that all the discussions on finite temperature properties near QCPs are not only applicable to the $p$-wave superfluid model in Sec. \ref{pwave}, but also to the TSC models in the following sections. These properties are universal and robust.

\section{TSC of Dirac fermions}\label{Dirac}
In the $p$-wave superfluid model discussed in Sec. \ref{pwave}, only QLMFA QCPs exist between gapped phase and NPP. 
More degrees of freedom (e.g., different orbitals) need to be introduced to realize all three classes of QLMF QCPs in a given model.
In this section, we discuss a TSC model of Dirac fermions with four bands labeled by spin and orbital degrees of freedom. In this model, all three classes of QLMF QCPs exist.
In the next section, we will show that similar physics also exist in the TSC model of Cu$_x$Bi$_2$Se$_3$.

Topological classifications of NPPs, NLPs and NSPs can be found in Refs. \cite{Zhao13,Kobayashi14, Zhao16}, and we refer readers to these references for general discussions. 
Generally speaking, the topological stability of these nodal phases in TSFs/TSCs further depends on additional global symmetries. We will discuss the symmetries that protect these nodal phases in concrete examples below.

In our discussions of QCPs, we assume the nodal phases are either protected by global symmetries and topology, or due to the absence of 
other gapping couplings in materials as a result of specific energetic non-topological reasons. 
We focus on the dynamics of QCPs between gapped TSCs and nodal phases assuming these phases are present and phase transitions do exist. 
The particular transitions we describe below offer concrete realizations of different QCPs in TSCs and detailed energetics 
of how to drive the corresponding transitions in terms of generalized mass fields.

Before considering pairing fields, we first introduce a low energy effective Hamiltonian for Dirac semimetals  \cite{Young12, Armitage18} with two orbitals
\begin{equation}
H_0=\sum_{\bf k}C^\dagger_{\bf k}\mathcal H_0({\bf k})C_{\bf k},
\end{equation}
where $C_{\bf k}=(c_{1,\uparrow,{\bf k}},c_{1,\downarrow,{\bf k}},c_{2,\uparrow,{\bf k}},c_{2,\downarrow,{\bf k}})^T$, 1 and 2 are orbital indices, and 
\begin{equation}
\mathcal{H}_0({\bf k})=v\sigma_z\otimes(s_xk_x+s_yk_y+s_zk_z).
\end{equation}
Here, $\sigma_\alpha$'s are Pauli matrices in orbital space. 
Let us rewrite the Hamiltonian in the Majorana representation. 
We define
\begin{gather}
\chi_{\bf k}=\begin{pmatrix}\chi_{1,{\bf k}}\\\chi_{2,{\bf k}}\end{pmatrix},
\end{gather}
where 
\begin{eqnarray}
\chi_{j,{\bf k}}=(\chi_{j,+,\uparrow}({\bf k}), \chi_{j,+,\downarrow}({\bf k}), \chi_{j,-,\uparrow}({\bf k}), \chi_{j,-,\downarrow}({\bf k}))^T,
\end{eqnarray}
and $j=1,2$.
Then the Hamiltonian of the Dirac semimetal becomes 
\begin{equation}
H_0=\frac12\sum_{\bf k}\chi^T_{-\bf k}\mathcal H_0^M({\bf k})\chi_{\bf k},
\end{equation}
where
\begin{equation}
\mathcal H_0^M({\bf k})=v\sigma_z\otimes (I\otimes s_xk_x-\tau_y\otimes s_yk_y+I\otimes s_zk_z).
\end{equation}

TSCs can be generated by introducing an odd-parity TRI intraorbital spin singlet pairing
\begin{equation}
\mathcal{H}_\Delta=\sigma_z\otimes\tau_x\otimes s_y\Delta.
\end{equation}
For convenience, we choose $\Delta>0$.
The Hamiltonian of Dirac TSCs is 
\begin{eqnarray}
&&H=\frac12\sum_{\bf k}\chi^T_{-\bf k}\mathcal{H}({\bf k})\chi_{\bf k}+H_I,\nonumber\\
&& \mathcal{H}({\bf k})=\mathcal H_0^M({\bf k})+\mathcal{H}_\Delta.
\end{eqnarray}
The interactions represented by $H_I$ are irrelevant operators in 3D.
Therefore, they are muted in the following discussions of QCPs.

The Hamiltonian is invariant under parity $\mathcal{P}=\sigma_x\otimes\tau_y$.
The bulk spectrum is fully gapped and isotropic.
This topological phase is protected by TRS.
Only TRB or sufficiently large TRI mass fields that commute with $\mathcal{H}_\Delta$ can drive quantum phase transitions.

\begin{table*}
\center
\begin{tabular}{|c|c|c|c|}
\hline
Description &Matrix Operator & Dirac TSCs& Cu$_x$Bi$_2$Se$_3$\\
\hline
{interorbital spin conserved hopping} & $\sigma_y\otimes I\otimes I$&Gapped$^*$&NPP\\
\hline
\multirow{2}{*}{interorbital spin exchange} & $\sigma_x\otimes\tau_y\otimes s_x$&Gapped$^*$&NPP\\
&  $-\sigma_x\otimes I\otimes  s_y$&Gapped$^*$&NPP\\
\hline
\multirow{3}{*}{\makecell{Zeeman-type\\intraorbital spin exchange}} & $I\otimes\tau_y\otimes s_x$&NPP&Gapped$^*$\\
&   $-I\otimes I\otimes  s_y$&NPP&Gapped$^*$\\
&  $I\otimes\tau_y\otimes s_z$&NPP&NPP\\
\hline
\multirow{3}{*}{\makecell{orbital dependent \\intraorbital spin exchange}}  & $\sigma_z\otimes\tau_y\otimes s_x$&NPP&NPP\\
&   $-\sigma_z\otimes I\otimes  s_y$&NPP&NPP\\
&  $\sigma_z\otimes\tau_y\otimes s_z$&NPP&Gapped$^*$\\
\hline
\end{tabular}
\caption{List of $U(1)$ symmetry invariant mass fields that lead to NPPs in Dirac TSC and/or Cu$_x$Bi$_2$Se$_3$ model. NPPs are realized when these fields are sufficiently large. All these fields break TRS. The interorbital spin conserved hopping and orbital dependent intraorbital spin exchange fields also break the existing parity symmetry of $\mathcal{H}({\bf k})$, while the rest do not. In some cases, the superconducting gap remain open regardless of the strength of these fields and we label these phases as `Gapped$^*$'. Here, the asterisk means the gap never closes and there is no phase transition.}\label{NPPtable}
\end{table*}

\begin{table*}
\center
\begin{tabular}{|c|c|c|c|}
\hline
Description& Matrix Operator & Dirac TSCs& Cu$_x$Bi$_2$Se$_3$\\
\hline
{orbital dependent chemical potential} & $\sigma_z\otimes\tau_y\otimes I$ &Gapped$^*$ &NLP \\
\hline
\multirow{3}{*}{interorbital spin exchange} & $\sigma_y\otimes I\otimes  s_x$ &NLP &NLP \\
& $-\sigma_y\otimes\tau_y\otimes s_y$ &NLP &NLP\\
 & $\sigma_y\otimes I\otimes s_z$ &NLP &Gapped$^*$\\
\hline
\end{tabular}
\caption{List of $U(1)$ symmetry invariant mass fields that lead to NLPs in Dirac TSC and/or Cu$_x$Bi$_2$Se$_3$ model. NLPs are realized when these fields are sufficiently large. All these fields are TRI but break the existing parity symmetry of $\mathcal{H}({\bf k})$. In some cases, the superconducting gap remain open regardless of the strength of these fields and we label these phases as `Gapped$^*$'. Here, the asterisk means the gap never closes and there is no phase transition.}\label{NLPtable}
\end{table*}

\begin{table*}
\center
\begin{tabular}{|c|c|c|c|}
\hline
Decription&Matrix Operator & Dirac TSCs& Cu$_x$Bi$_2$Se$_3$\\
\hline
{interorbital spin singlet pairing}& $\sigma_x\otimes\tau_z\otimes s_y$&NSP&NSP\\ 
\hline
intraorbital spin singlet pairing& $\sigma_z\otimes\tau_z\otimes s_y$&Gapped$^*$ &NPP\\
\hline
\multirow{3}{*}{interorbital spin triplet pairing} & $\sigma_y\otimes\tau_z\otimes I$&NPP&NPP\\
&  $\sigma_y\otimes\tau_x\otimes s_z$&NPP&NPP\\
&  $\sigma_y\otimes\tau_x\otimes s_x$&NPP&Gapped$^*$\\
\hline
\end{tabular}
\caption{List of $U(1)$ symmetry breaking mass fields that lead to NSPs and NPPs in Dirac TSC and/or Cu$_x$Bi$_2$Se$_3$ model. Nodal phases are realized when these terms are sufficiently large. All these terms break TRS. The interorbital spin singlet pairing field breaks the existing parity symmetry of $\mathcal{H}({\bf k})$, while the rest do not. In some cases, the superconducting gap remain open regardless of the strength of these fields and we label these phases as `Gapped$^*$'. Here, the asterisk means the gap never closes and there is no phase transition.}\label{pairing}
\end{table*}

\begin{table*}
\center
\begin{tabular}{|c|c|c|c|}
\hline
Description & Matrix Operator & Dirac TSCs& Cu$_x$Bi$_2$Se$_3$\\
\hline
intraorbital spin singlet pairing  & $I\otimes\tau_x\otimes s_y$ & Gapped & Gapped\\
\hline
interorbital spin conserved hopping  & $\sigma_x\otimes\tau_y\otimes I$ & Gapped & Gapped\\
\hline
\end{tabular}
\caption{List of mass fields that lead to transitions to different gapped phases when these mass fields are sufficiently large. All these fields are TRI. The intraorbital spin singlet pairing breaks the existing parity symmetry of $\mathcal{H}({\bf k})$, while the interorbital spin conserved hopping does not. }\label{gap}
\end{table*}

\subsection{Quantum criticality in the bulk}\label{DiracEFT}
This TSC model of Dirac fermions can host all three classes of  QLMF QCPs and all three types of nodal superconducting phases.
In the following, we show explicitly which mass fields will lead to these nodal phases and corresponding QCPs.

We first consider phase transitions driven by $U(1)$ invariant non-pairing mass fields. In this case, the order parameter does not change across the transitions.
It is worth noting that weak magnetic fields cannot penetrate into the bulk of superconductors due to Meissner effect \cite{SCbook}. Our discussions below mainly apply to the effects of various internal spin-exchange fields ${\bf J}$.
We find that both NPPs and NLPs exist as a result of spin exchange fields.
In particular, NPPs only exist if TRS is broken; while NLPs only exist if TRS is preserved (see Tables \ref{NPPtable} and \ref{NLPtable}).

To realize NSP in this model, it is necessary to introduce additional pairing fields that also break $U(1)$ gauge symmetry.
In fact, introducing additional pairing fields can lead to both NSPs and NPPs  when TRS is broken (see Table \ref{pairing}).

In addition to nodal phases, additional mass fields (pairing or non-pairing) can also lead to phase transitions to different gapped phases when these mass fields are large enough (see Table \ref{gap}). 
For example, the TRI  even-parity intraorbital spin singlet pairing $I\otimes\tau_x\otimes s_y\Delta'$ drives a transition to a non-topological superconducting phase when $\Delta'>\Delta$. 
The TRI interorbital hopping $\sigma_x\otimes\tau_y\otimes It$ drives a transitions to a topological insulating phase when $t>\Delta$.

Following the arguments in Sec. \ref{TQCP}, we find that all the phase transitions discussed in this section only exist at zero temperature and corresponds to topological QCPs.

For other mass fields allowed by charge conjugation symmetry but not listed in the Tables, the superconducting gap remains open regardless of the magnitude of these fields. Thus, these mass fields do not lead to phase transitions, and we will not discuss them in details.

\subsubsection{QLMFA and NPPs.}
TRB intraorbital spin exchange fields can lead to NPPs.
We list them in Table \ref{NPPtable}.
The bulk spectrum near phase transitions in the presence of these fields becomes
\begin{equation}
E_{\bf k}=\sqrt{v^2{\bf k}^2_\perp+(\sqrt{v^2k_\parallel^2+\Delta^2}\pm J)^2},
\end{equation}
which has two point nodes at ${\bf k}_\perp=0$ and $k_\parallel=\pm\sqrt{J^2-\Delta^2}/v$ when $J>\Delta$.
$k_\parallel$ (${\bf k}_\perp$) is the momentum parallel (perpendicular) to the spin exchange field ${\bf J}$, and $J=|{\bf J}|$.

The NPPs are SPT states.
We take the orbital dependent spin exchange field $\sigma_z\otimes \tau_y\otimes s_zJ$ as an example. In this case, both $\mathcal T$ and $\mathcal P$ are broken, but the Hamiltonian has a combined $\mathcal{TP}$ symmetry. In addition, there also exists a reflection symmetry across $z$-axis $M_z=\sigma_z\otimes\tau_y\otimes s_z$, under which the Hamiltonian transforms as
\begin{equation}\label{Mz}
M_z\mathcal{H}({\bf k})M_z^{-1}=\mathcal{H}(-k_x,-k_y,k_z).
\end{equation}
The point nodes are protected by both $\mathcal{TP}$ and $M_z$.

We can construct the effective Hamiltonian  at low energy using similar procedures as described in Sec. \ref{EFT}.
The effective Hamiltonian near $\Delta=J$ can be written as
\begin{multline}\label{mzeff}
H_\text{eff}=\frac12\sum_{\bf k}\chi^T_{\bf -k}\Bigg[\Gamma_z\left(\Delta-J+\frac{v^2k_z^2}{2\Delta}\right)\\
+\Gamma_xvk_x+\Gamma_yvk_y\Bigg]\chi_{\bf k},
\end{multline}
where 
\begin{align}
&&\Gamma_x=P(\sigma_z\otimes I\otimes s_x)P, && \Gamma_y=P(-\sigma_z\otimes \tau_y\otimes s_y)P, \nonumber\\
&&\Gamma_z=P(\sigma_z\otimes\tau_x\otimes s_y)P,
\end{align}
and
\begin{equation}
P=P^\tau_{z,+}P^s_{x,-}+P^\tau_{z,-}P^s_{x,+}.
\end{equation}
The effective Hamiltonian belongs to QLMFA class.
These phase transitions are $7/2$th order in 3D.

We notice that Eq. (\ref{Mz}) implies an emergent parity symmetry $\mathcal P$ in the effective Hamiltonian (\ref{mzeff}), which is absent in the original Hamiltonian with $J$ included.

Similarly, TRB interorbital spin triplet pairing fields listed in Table \ref{pairing} also lead to QLMFA QCPs.

\subsubsection{QLMFB and NLPs.}
TRI spin exchange fields can lead to  NLPs. We list them in Table \ref{NLPtable}.
The bulk spectrum near transitions taken into account of these fields is
\begin{equation}
E_{\bf k}=\sqrt{v^2k^2_\parallel+(\sqrt{v^2{\bf k}^2_\perp+\Delta^2}\pm J)^2},
\end{equation}
which has line nodes at $k_\parallel=0$ and ${\bf k}_\perp^2=(B^2-\Delta^2)/v^2$ when $J>\Delta$.

The NLPs are SPT states.
For example, for $-\sigma_y\otimes\tau_y\otimes s_yJ$, $\mathcal P$ symmetry is broken. The NLP is protected by TRS and mirror reflection with respect to $xz$-plane $M_{xz}=\sigma_y\otimes\tau_y\otimes s_y$, under which the Hamiltonian transforms as
\begin{equation}\label{Mxz}
M_{xz}\mathcal{H}({\bf k})M_{xz}^{-1}=\mathcal{H}(k_x,-k_y,k_z).
\end{equation}
The effective Hamiltonian can be constructed similarly
\begin{multline}\label{mxzeff}
H_\text{eff}=\frac12\sum_{\bf k}\chi^T_{\bf -k}\Bigg[\Gamma_x\left(\frac{v^2}{2\Delta}(k_x^2+k_z^2)+\Delta-J\right)\\
+\Gamma_yvk_y\Bigg]\chi_{\bf k},
\end{multline}
where
\begin{align}
&&\Gamma_x=P(\sigma_z\otimes\tau_x\otimes s_y)P, && \Gamma_y=P(-\sigma_z\otimes\tau_y\otimes s_y)P,
\end{align}
and
\begin{equation}
P=P^\sigma_{x,+}P^\tau_{z,+}+P^\sigma_{x,-}P^\tau_{z,-}.
\end{equation}
Here $P^\sigma_{\alpha,\pm}=(1\pm\sigma_\alpha)/2$.
The effective Hamiltonian belongs to QLMFB universality class.
The phase transitions are 3rd order in 3D.

We again note that Eq. (\ref{Mxz}) implies an emergent parity symmetry $\mathcal P$ in the effective Hamiltonian (\ref{mxzeff}), which is absent in the original Hamiltonian with $J$ included.

\subsubsection{QLMFC and NSPs.}
With the TRB interorbital spin singlet pairing $\sigma_x\otimes\tau_z\otimes s_yD$, the bulk spectrum near transitions is
\begin{equation}
E_{\bf k}=\left|\sqrt{v^2k^2+\Delta^2}\pm D\right|,
\end{equation}
which has surface nodes at $k^2=(D^2-\Delta^2)/v^2$ when $D>\Delta$.

Nodal surface states are generally less stable. Here we simply treat the QCP as a multicritical point.
The effective Hamiltonian
\begin{equation}
H_\text{eff}=\frac12\sum_{\bf k}\chi^T_{\bf -k}\Gamma\left(\frac{v^2k^2}{2\Delta}+\Delta-D\right)\chi_{\bf k},
\end{equation}
 belongs to QLMFC class with 
\begin{equation}
\Gamma=P(\sigma_z\otimes\tau_x\otimes s_y)P
\end{equation}
and
\begin{equation}
P=P^\sigma_{y,+}P^\tau_{y,-}+P^\sigma_{y,-}P^\tau_{y,+}.
\end{equation}
These phase transitions are $5/2$th order in 3D.

\subsection{Surface quantum criticality}
Gapless helical Majorana states exist on the surfaces of TRI Dirac TSCs.
In the presence of Zeeman-type or orbital dependent intraorbital spin exchange fields that lead to NPPs, the surface states can be gapped by the field $J_\perp$ perpendicular to this surface. Topological phase transitions happen on the surface between two gapped  states when $J_\perp$ is tuned across zero. On a given surface, $J_\perp=0$, $T=0$ corresponds to a surface QCP.

\section{Topological superconducting C\lowercase{u$_x$}B\lowercase{i}$_2$S\lowercase{e}$_3$ model}\label{CBS}
As another example, we discuss the QLMF QCPs in the TSC Cu{$_x$}Bi$_2$Se$_3$ model.
We first tune the topological insulating gap and chemical potential to zero to obtain a semimetal Hamiltonian \cite{Zhang09}, which can be written in the Majorana representation at low energy as
\begin{equation}
\mathcal{H}'_0{}^M({\bf k})=v\sigma_z\otimes(\tau_y\otimes s_y k_x+I\otimes s_x k_y)+v_z\sigma_y\otimes\tau_y\otimes I k_z,
\end{equation}
$v\neq v_z$ due to the crystal symmetry.

Following the criterion given by Fu and Berg \cite{Fu10}, the TRI odd-parity interorbital spin triplet pairing 
\begin{equation}
\mathcal{H}'_\Delta=\sigma_y\otimes\tau_z\otimes s_x\Delta
\end{equation}
should generate a fully gapped TSC.
The TSC Hamiltonian 
\begin{equation}
H=\frac12\sum_{\bf k}\chi^T_{-\bf k}(\mathcal H'_0{}^M({\bf k})+\mathcal{H}'_\Delta)\chi_{\bf k}+H_I,
\end{equation}
is invariant under parity transformation $\mathcal{P}=\sigma_x\otimes\tau_y$. The interactions in $H_I$ are irrelevant operators and will be muted for the discussions of QCPs.
The topological phase is protected by TRS.

\subsection{QCPs in the bulk}
In this model, we also have all three classes of QLMF QCPs.
Similar to the Dirac TSC model, NPPs and NLPs can be generated by $U(1)$ invariant non-pairing mass fields. NPPs exist when TRS is broken; NLPs exist when TRS is preserved. NSP must be generated by an additional TRB pairing field.
We list all mass fields that lead to nodal phases in Tables \ref{NPPtable}, \ref{NLPtable} and \ref{pairing}.
These nodal phases are protected by relevant symmetries depending on the operators driving the transitions. We do not discuss each case individually.
The QCPs associated with transitions to these nodal phases belong to their corresponding QLMF universality classes.
The low energy effective field theory near QCPs can be obtain similarly as in Sec. \ref{DiracEFT}, and we do not list them here.

\subsection{Surface QCPs}
Surface QCPs also exist in TSC Cu$_x$Bi$_2$Si$_3$. 
The gapless Majorana surface states in TSC Cu$_x$Bi$_2$Si$_3$  can be gapped by some TRB mass fields that lead to NPPs. The mass fields that open gaps on the surface depend on the orientation of these surfaces.
For example, Zeeman-type spin exchange field along $z$-direction $I\otimes\tau_y\otimes s_zJ$ and TRB interorbital hopping $\sigma_y\otimes I\otimes It$ can open gaps on surfaces perpendicular to $z$-axis with any finite strength.
Surface states are quantum critical when these fields are zero.

\section{Conclusions}
In conclusion, we have investigated a broad set of QCPs in TSFs and TSCs. These QCPs define quantum phase transitions driven by generalized mass fields between fully gapped TSFs/TSCs and nodal phases. Phases on two sides of the transition can break the same symmetries and have the same local ordering but with different global topologies.
These QCPs therefore are beyond the standard Landau paradigm of order-disorder phase transitions. U(1) symmetry is also spontaneously broken at these QCPs.

We have identified three main universality classes that have distinct scaling properties and are
characterized by generalized QLMFs. The main conclusions are:

(1) All the QCPs studied here can naturally emerge when generalized Zeeman (spin
exchange) fields or other relevant fields are varied. QCPs separate states with
different global topologies. The upper critical dimensions of these QCPs are either $D_u=3/2$ or $D_u=2$
depending the classes QLMFA, QLMFB, QLMFC, to which they belong. Below or at the
upper critical dimension, QCPs are described by strong coupling conformal field theory
fixed points; while above it free quantum Lifshitz Majorana fermions are robustly stable.

(2) These QCPs induce various subtle non-analytical cusp structures in bulk quantities such as generalized susceptibility. Each QLMF class has its own unique bulk
signatures as a smoking gun of topological QCPs. For instance, in 3D the non-analytical
structures associated with QCPs are of 7/2th, 3rd and 5/2th order for QLMFA, QLMFB, and QLMFC, respectively.
They are generally smoother than a typical 2nd order phase transition.

(3) For transitions driven by the generalized mass fields that lead to QLMFA QCPs, as
precursors to transitions in the bulk, surface states can be gapped by arbitrarily small fields perpendicular to the surface. This critical behavior of surface states, i.e., the surface states being
quantum critical at zero field, leads to non-analytical surface spin susceptibilities that
can be potentially studied in experiments.  In fact, the susceptibility itself, being an
even function of the field, has a non-analytical part that is proportional to the magnitude of Zeeman fields, indicating a
cusp structure.

(4) There exist no finite temperature transitions between the phases we have studied.
All the cusp structures disappear once temperatures become finite and non-analytical
structures are replaced with smooth crossovers. The physics in the quantum critical regime
is completely defined by the quantum criticality physics, and the temperature scaling of thermodynamic quantities are distinct for each class of QLMFs. This can be potentially important, as in practical situations it is likely the distinct temperature scaling dictated by QCPs in quantum
critical regimes rather than the $T=0$ cusp structures that can be measured.

(5) In a few concrete models such as $p$-wave superfluids, and TSCs of Dirac fermions and Cu$_x$Bi$_2$Se$_3$,
we have found detailed realizations of the QCPs and universality classes discussed above. These
concrete studies are intended to bring the physics of QCPs one step closer to physical
reality.

There are a few very exciting issues we plan to explore in the near future. The first one is related to the
strong coupling conformal field theory (CFT) fixed point in (2+1)D. As we have stated in the article, although
there may be no clear distinctions between a free field theory QCP and a QCP of a CFT fixed
point, the transport properties and hydrodynamics in these two classes of QCPs should be very different. It
remains to be understood the transport properties and hydrodynamics near a QCP of a CFT fixed point, which we speculate to be highly universal as well.
The second issue is perhaps the relation between Gross-Neveu strong coupling fixed point in
(1+1)D relativistic theory \cite{Gross74}, and the relevant interactions in (2+1)D QLMFB/QLMFC implied by the
scaling argument or a simple 1-loop calculation.  [In (2+1)D, QLMFB and QLMFC are identical.] It is possible that QLMFB/QLMFC presents a
generalization of the Gross-Neveu CFT field but in two spatial dimensions. If this is true,
QLMFB/QLMFC maybe a new candidate for CFT but in (2+1)D. It remains to be investigated in the
future, perhaps in the context of large $N$-expansion.

\begin{acknowledgments}
We would like to thank Ian Affleck, Zheng-Cheng Gu, Hae-Young Kee, Sung-Sik Lee, and Xiao-Gang Wen for helpful discussions. This work is partially support by Canadian Institute for Advanced Research. F. Y. is supported by a four year doctoral fellowship from UBC. F. Z. would also
like to acknowledge the hospitality of the 2020 CUHK winter workshop on Quantum criticality and topological phases, during which many fascinating gapless topological states were highlighted and debated.
\end{acknowledgments}

\appendix

\section{Quantum phase transitions in superfluids with competing $s$- and $p$-wave pairings}\label{sp}

In Sec. \ref{pwavemodel}, we mentioned that in TSFs if symmetry allows competing $s$- and $p$-wave pairings, there can be phase transitions or crossovers between topological and non-topological superfluids.
Here, we discuss them in details.

We choose the same TRI $p$-wave pairing $\Delta_p({\bf k})$.
If the $s$-wave pairing is also TRI, we expect the topological phase to persist for weak $s$-wave pairing, because the topological phase is protected by TRS. On the other hand, pure $s$-wave superfluids are always topologically trivial. Therefore, we expect a topological phase transition into this non-topological phase when the $s$-wave pairing becomes strong enough.  

On the other hand, if the $s$-wave pairing breaks TRS, the non-trivial topology is no longer protected and could be broken immediately by any finite TRB pairing.
As shown below, the superconducting gap never closes in this case. Instead of phase transitions, there exist a crossover between the topological and non-topological states.

Let us introduce an $s$-wave order parameter $\Delta_s=|\Delta_s|e^{i\theta}=\Delta_s^I+i\Delta_s^B$, with phase $\theta$ relative to $\Delta_p({\bf k})$. 
Here, $\Delta_s^I=|\Delta_s|\cos\theta$ ($\Delta_s^B=|\Delta_s|\sin\theta$) is the TRI (TRB) $s$-wave pairing.
In the Majorana representation, the $s$-wave pairing field can be written as 
\begin{equation}
H_s=\frac12\sum_{\bf k}\chi^T_{ -\bf k}(\tau_x\otimes s_y\Delta_s^I-\tau_z\otimes s_y\Delta_s^B)\chi_{\bf k}.
\end{equation}

\subsection{TRI $s$-wave pairing}
For $\theta=0$ or $\pi$, we get the TRI $s$-wave pairing $\tau_x\otimes s_y \Delta_s^I$. 
The energy spectrum for the quasiparticles is
\begin{equation}\label{spectrum}
E_{\bf k}^{(\pm)}=\sqrt{(vk\pm |\Delta_s^I|)^2+(\epsilon_k-\mu)^2}.
\end{equation}
Notice that we cannot drop the $\epsilon_k$ term even it is of higher order of ${\bf k}$. This is because the minima of the bulk spectrum are at finite ${\bf k}$ near phase transitions.
For $\mu>0$, we expect topological phase transitions as the ratio between $s$- and $p$-wave pairings $|\Delta_s^I|/v$ is varied.
Indeed, when $|\Delta_s^I|=v\sqrt{2\mu}$, the bulk gap closes on a surface given by $k=\sqrt{2\mu}$.
Phase transitions occur between two gapped phases: topological superfluids for $|\Delta_s^I|<v\sqrt{2\mu}$ and non-topological superfuids for $|\Delta_s^I|>v\sqrt{2\mu}$.

It is worth noting that these phase transitions, albeit between gapped phases, do not belong to the universality class of Lorentz invariant Majorana fields studied in Ref. \cite{Yang19}, since the bulk gap closes on a surface $k=\sqrt{2\mu}$ rather than at ${\bf k}=0$.

\subsection{TRB $s$-wave pairing}
For $\theta=\pm\pi/2$, the $s$-wave pairing $-\tau_z\otimes s_y \Delta_s^B$ breaks TRS, and the bulk spectrum 
\begin{equation}
E_{\bf k}=\sqrt{v^2k^2+(\Delta_s^B)^2+(\epsilon_k-\mu)^2}
\end{equation}
is always gapped except at $\mu=\Delta_s^B=0$. There exists a tricritical point at $\mu=\Delta_s^B=0$.
Rather than phase transitions, the topological and non-topological states can be smoothly connected by a crossover.
In contrast to the TRI $s$-wave  pairing case, here we can neglect the $\epsilon_k$ term when studying universality, as the bulk gap can only close at ${\bf k}=0$. 
In the low energy limit, the tricritical point is described by Lorentz invariant free Majarana fields with two anticommuting mass terms $\mu$ and $\Delta_s^B$.

Any finite TRB $s$-wave pairing immediately opens a gap on all surfaces.
For example, on the surface $y=0$, the surface Hamiltonian is
\begin{equation}\label{TRBsurface}
H_\text{surf}=\frac12\sum_{\bf k}\psi^T_{-{\bf k},y}(- s_zvk_x+ s_xvk_z- s_y \Delta_s^B)\psi_{{\bf k},y}.
\end{equation}
Hamiltonians on other surfaces have similar forms.

Topological QCPs also exist on surfaces. Surface Hamiltonians are quantum critical at $\Delta_s^B=0$.
When the TRB $s$-wave pairing $\Delta_s^B$ is tuned across zero, topological quantum phase transitions happens on all surfaces simultaneously between two gapped surface phases with different topologies. The Chern number associated with the surface Hamiltonian changes by one across the transition. These QCPs are described by Lorentz invariant free Majorana fields and the zero temperature transitions are 3rd order.

In 2D, $s$-wave pairings have similar effects.

\section{Phase transitions in $p$-wave superfluids driven by Zeeman field in the weak coupling limit}\label{weaktransitions}
Here we discuss the topological quantum phase transitions between a fully gapped phase and an NPP in the weak coupling $p$-wave superfluid model.
We present the low energy Hamiltonian in the momentum space near phase transitions.

In the weak coupling limit $v^2\ll\mu$, the chemical potential is approximately at the Fermi energy $\mu\approx\epsilon_F>0$. The transition hapens at $B=B_c\approx vk_F$ when the gap closes at ${\bf k}_\perp=0$, $k_\parallel =\pm K\approx\pm k_F$, $k_F$ being the Fermi momentum and $K=\sqrt{2(\mu-v^2)}$. 
Without loss of generality, we choose the magnetic field to be along $y$-axis.

There are two point nodes in the spectrum at critical point. Let us first write the Hamiltonian in the low energy limit near one of them, $k_x= k_z=0$, $k_y=-K\approx -k_F$. 
We first project the Hamiltonian onto the low energy subspace using projection operator
\begin{equation}
P^{(-)}=P^\tau_{x,+}P^s_{y,-}+P^\tau_{x,-}P^s_{y,+}.
\end{equation}
The projected Hamiltonian is
\begin{equation}\label{proj}
\mathcal{H}^{(-)}_\text{proj}({\bf k})=-\Gamma_y^{(-)}(vk_y+B_y)+\Gamma_x^{(-)}vk_x+\Gamma_z^{(-)}vk_z.
\end{equation}
where 
\begin{align}\label{weakGamma}
\Gamma_x^{(-)}=P^{(-)}(-\tau_z\otimes s_z) P^{(-)},\nonumber\\
\Gamma_z^{(-)}=P^{(-)}(\tau_z\otimes s_x) P^{(-)},\nonumber\\
\Gamma_y^{(-)}=P^{(-)}(I\otimes s_y)P^{(-)}.
\end{align}
In this particular limit of weak coupling, Eq. (\ref{proj}) obtained by a simple projection suggests a Lorentz symmetry near the point node.
This emergent Lorentz symmetry is an artifact of projection that  turns out to be inadequate for scaling in this case.  
We need to further take into account the couplings between the low energy states near point nodes and high energy states.

By integrating out the high energy degrees of freedom, we obtain the leading order contribution from the coupling
\begin{equation}
\mathcal{H}^{(2)}({\bf k})=-\frac{(\mu-\epsilon_k)^2}{2vk_y}\Gamma_y^{(-)}.
\end{equation}
Combining it with the projected Hamiltonian (\ref{proj}) and expanding $k_y$ near the point node $k_y=\delta k_y-K$, we obtain the Hamiltonian in the low energy limit near this point node
\begin{multline}
\mathcal{H}^{(-)}_\text{eff}({\bf k})=\Gamma_y^{(-)}\Big[(B_c- B_y)+\frac{\mu-v^2}{B_c}(\delta k_y)^2\Big]\\
+\Gamma_x^{(-)}vk_x+\Gamma_z^{(-)}vk_z.
\end{multline}

Similarly, near the other point node $k_x= k_z=0$, $k_y=K\approx k_F$. The Hamiltonian in the low energy limit is
\begin{multline}
\mathcal{H}^{(+)}_\text{eff}({\bf k})=\Gamma_y^{(+)}\Big[(B_c- B_y)+\frac{\mu-v^2}{B_c}(\delta k_y)^2\Big]\\
+\Gamma_x^{(+)}vk_x+\Gamma_z^{(+)}vk_z,
\end{multline}
where $\Gamma^{(+)}_\alpha$'s are obtained by replacing $P^{(-)}$ with 
\begin{equation}
P^{(+)}=P^\tau_{x,+}P^s_{y,+}+P^\tau_{x,-}P^s_{y,-},
\end{equation}
in Eq. (\ref{weakGamma}). Notice that $P^{(-)}+P^{(+)}=I\otimes I$. Near this point node, we have $k_y=\delta k_y+K$.

The Hamiltonian involving both point nodes is a $4\times4$ matrix in the low energy limit
\begin{equation}
\mathcal{H}_\text{eff}({\bf k})=\mathcal{H}^{(-)}_\text{eff}({\bf k})\oplus\mathcal{H}^{(+)}_\text{eff}({\bf k}).
\end{equation}
This Hamiltonian is valid in the low energy regime $E_{\bf k}^{(-)}\ll v^2=\frac{|\Delta_p(k=k_F)|^2}{2\epsilon_{F}}$.
The NPP is protected by parity symmetry $\mathcal P=\tau_y$.


\begin{thebibliography} {References} 
\bibitem{Volovik88} G. E. Volovik, J. Exp. Theor. Phys. {\bf67}, 1804 (1988) [Russian original: Zh. Eksp. Teor. Fiz. {\bf94}, 123 (1988)]. 

\bibitem{Volovik} G. E. Volovik, {\it The Universe in a Helium Droplet} (Oxford University Press, 2003).

\bibitem{Read00} N. Read and D. Green, Phys. Rev. B, {\bf61}, 10267 (2000).

\bibitem{Moore91} G. Moore and N. Read, Nucl. Phys. B {\bf360}, 362 (1991). 

\bibitem{Nayak96} C. Nayak and F. Wilczek, Nucl. Phys. B {\bf479}, 529 (1996). 

\bibitem{Read96} N. Read and E. Rezayi, Phys. Rev. B {\bf54}, 16864 (1996).

\bibitem{Ivanov01} D. A. Ivanov, Phys. Rev. Lett. {\bf86}, 268 (2001).

\bibitem{Kitaev01} A. Kitaev, Phys.-Uspekhi {\bf44}, 131 (2001).

\bibitem{Kitaev03} A. Kitaev, Ann. Phys. {\bf303}, 2 (2003).

\bibitem{Roy08} R. Roy, arXiv: 0803.2868.

\bibitem{Qi09} X.-L. Qi, T. L. Hughes, S. Raghu, and S.-C. Zhang, Phys. Rev. Lett. {\bf 102}, 187001 (2009).

\bibitem{Qi11} X.-L. Qi and S.-C. Zhang, Rev. Mod. Phys. {\bf83}, 1057 (2011).

\bibitem{Bernevig} B. A. Bernevig and T. L. Hughes, {\it Topological insulators and topological superconductors} (Princeton University Press, 2013).

\bibitem{Zhang13} F. Zhang, C. L. Kane, and E. J. Mele, Phys. Rev. Lett. {\bf111}, 056402 (2013).

\bibitem{Mizushima16} T. Mizushima, Y. Tsutsumi, T. Kawakami, M. Sato, M. Ichioka, and K. Machida, J. Phys. Soc. Jpn. {\bf85}, 022001 (2016).

\bibitem{Sato17} M. Sato and Y. Ando, Rep. Prog. Phys. {\bf 80}, 076501 (2017).

\bibitem{Kane05} C. L. Kane and E. J. Mele, Phys. Rev. Lett. {\bf95} 146802 (2005); Phys. Rev. Lett. {\bf 95}, 226801 (2005).

\bibitem{Bernevig06a} B. A. Bernevig and Shou-Cheng Zhang, Phys. Rev. Lett. 96, 106802 (2006).

\bibitem{Bernevig06b} B. A. Bernevig, T. L. Hughes, and S. C. Zhang, Science {\bf314}, 1757 (2006).

\bibitem{Fu07} L. Fu, C. L. Kane and E. J. Mele, Phys. Rev. Lett. {\bf98}, 106803 (2007). 

\bibitem{Moore07} J. E. Moore and L. Balents, Phys. Rev. B {\bf75} 121306(R) (2007).

\bibitem{Fu07b} L. Fu and C. L. Kane, Phys. Rev. B {\bf76}, 045302 (2007).

\bibitem{Qi08} X. L. Qi, T. L. Hughes, and S.-C. Zhang, Phys. Rev. B {\bf78}, 195424 (2008).

\bibitem{Hasan10} M. Z. Hasan and C. L. Kane, Rev. Mod. Phys. {\bf82}, 3045 (2010).

\bibitem{Fu08} L. Fu and C. L. Kane, Phys. Rev. Lett. {\bf100}, 096407 (2008).

\bibitem{Qi10a} X.-L. Qi, T. L. Hughes, and S.-C. Zhang, Phys. Rev B {\bf82}, 184516 (2010).

\bibitem{Lutchyn10} R. M. Lutchyn, J. D. Sau, and S. Das Sarma, Phys. Rev. Lett. {\bf105}, 077001 (2010).

\bibitem{Chung11} S. B. Chung, H.-J. Zhang, X.-L. Qi, and S.-C. Zhang, Phys. Rev. B 84, 060510(R) (2011).

\bibitem{Nakosai12} S. Nakosai, Y. Tanaka, and N. Nagaosa, Phys. Rev. Lett. {\bf108}, 147003 (2012).

\bibitem{Schnyder08} A. P. Schnyder, S. Ryu, A. Furusaki, and A. W. W. Ludwig, Phys. Rev. B {\bf 78}, 195125 (2008).

\bibitem{Kitaev09} A. Kitaev, AIP Conference Proceedings {\bf1134}, 22 (2009)

\bibitem{Qi10} X.-L. Qi, T. L. Hughes, and S.-C. Zhang, Phys. Rev. B {\bf 81}, 134508 (2010).

\bibitem{Teo10} J. C. Y. Teo and C. L. Kane, Phys. Rev. B {\bf 82}, 115120 (2010).

\bibitem{Sato06} M. Sato, Phys. Rev. B {\bf73}, 214502 (2006).

\bibitem{Beri10} B. B{\'e}ri, Phys. Rev. B {\bf81}, 134515 (2010).

\bibitem{Zhao13} Y. X. Zhao and Z. D. Wang, Phys. Rev. Lett. {\bf110}, 240404 (2013).

\bibitem{Kobayashi14} S. Kobayashi, K. Shiozaki, Y. Tanaka, and M. Sato, Phys. Rev. B {\bf90}, 024516 (2014).

\bibitem{Zhao16} Y. X. Zhao, A. P. Schnyder, and Z. D. Wang, Phys. Rev. Lett. {\bf116}, 156402 (2016).

\bibitem{Wan11} X. Wan, A. M. Turner, A. Vishwanath, and S. Y. Savrasov, Phys. Rev. B 83, 205101 (2011).

\bibitem{Burkov11} A. A. Burkov and L. Balents, Phys. Rev. Lett. 107, 127205 (2011).

\bibitem{Burkov18} A. A. Burkov, Annu. Rev. Condens. Matter Phys. 9, 359 (2018).

\bibitem{Armitage18} N. P. Armitage, E. J. Mele, and A. Vishwanath, Rev. Mod. Phys. {\bf 90}, 015001 (2018).

\bibitem{Meng12} T. Meng and L. Balents, Phys. Rev. B {\bf86}, 054504 (2012); Phys. Rev. B {\bf 96}, 019901(E) (2017).

\bibitem{Cho12} G. Y. Cho, J. H. Bardarson, Y.-M. Lu, and J. E. Moore, Phys. Rev. B {\bf86}, 214514 (2012).

\bibitem{Yang14} S. A. Yang, H. Pan, and F. Zhang, Phys. Rev. Lett. {\bf113}, 046401 (2014).

\bibitem{Bednik15} G. Bednik, A. A. Zyuzin, and A. A. Burkov, Phys. Rev. B 92, 035153 (2015).

\bibitem{Yada11} K. Yada, M. Sato, Y. Tanaka, and T. Yokoyama, Phys. Rev. B {\bf83}, 064505 (2011).

\bibitem{Sato11} M. Sato, Y. Tanaka, K. Yada, and T. Yokoyama, Phys. Rev. B {\bf83}, 224511 (2011).

\bibitem{Schnyder11} A. P. Schnyder and S. Ryu, Phys. Rev. B {\bf84}, 060504(R) (2011).

\bibitem{Schnyder12} A. P. Schnyder, P. M. R. Brydon, and C. Timm, Phys. Rev. B {\bf 85}, 024522 (2012).

\bibitem{Agterberg17} D. F. Agterberg, P. M. R. Brydon, and C. Timm, Phys. Rev. Lett. {\bf118}, 127001 (2017).

\bibitem{Landau} L. D. Landau and E. M. Lifshitz, {\it Statistical Physics Part 1, 3rd Edition} (Butterworth-Heinemann, 1980).

\bibitem{Sachdev} S. Sachdev, Quantum Phase Transitions (Cambridge University Press, 1999).

\bibitem{Lifshitz41} E. M. Lifshitz, Zh. Eksp. Teor. Fiz. {\bf 11}, 255 (1941); Zh. Eksp. Teor. Fiz. {\bf 11}, 269 (1941).

\bibitem{Yang19} F. Yang, S.-J. Jiang, and F. Zhou, Phys. Rev. B {\bf100}, 054508 (2019).

\bibitem{BJYang14} B.-J. Yang, E.-G. Moon, H. Isobe, and N. Nagaosa, Nat. Phys. {\bf 10}, 774 (2014).

\bibitem{Cho16} G. Y. Cho and E.-G. Moon, Sci. Rep. {\bf 6} 19198 (2016).

\bibitem{Isobe16} H. Isobe, B.-J. Yang, A. Chubukov, J. Schmalian, and N. Nagaosa, Phys. Rev. Lett. {\bf 116}, 076803 (2016).

\bibitem{Senthil04a} T. Senthil, L. Balents, S. Sachdev, A. Vishwanath, and M. P. A. Fisher, Phys. Rev. B 70, 144407 (2004).

\bibitem{Senthil04b} T. Senthil, A. Vishwanath, Leon Balents, S. Sachdev, and M. P. A. Fisher, Science 303, 1490 (2004).

\bibitem{Rokhsar88}D. S. Rokhsar and S. A. Kivelson, Phys. Rev. Lett. {\bf61},  2376 (1988).

\bibitem{Haldane88} F. D. M. Haldane, Phys. Rev. Lett. {\bf61}, 1029 (1988).

\bibitem{Read91} N. Read and S. Sachdev, Phys. Rev. Lett. {\bf66}, 1773 (1991).

\bibitem{Moessner01} R. Moessner, and S. L. Sondhi, Phys. Rev. Lett. {\bf86}, 1881 (2001).

\bibitem{Fu10} L. Fu and E. Berg, Phys. Rev. Lett. {\bf 105}, 097001 (2010).

\bibitem{Sasaki11} S. Sasaki, M. Kriener, K. Segawa, K. Yada, Y. Tanaka, M. Sato, and Y. Ando, Phys. Rev. Lett. {\bf 107}, 217001 (2011).

\bibitem{Chen10}  X. Chen, Z.-C. Gu, and X.-G. Wen, Phys. Rev. B {\bf 82}, 155138 (2010); Phys. Rev. B {\bf 84}, 235128 (2011).

\bibitem{Chen12} X. Chen, Z.-C. Gu, Z.-X. Liu, and X.-G. Wen, Science {\bf 338}, 1604 (2012).

\bibitem{Chen13} X. Chen, Z.-C. Gu, Z.-X. Liu, and X.-G. Wen, Phys. Rev. B 87, 155114 (2013).

\bibitem{Lifshitz60} I. M. Lifshitz, J. Exp. Theor. Phys. {\bf 11}, 1130 (1960) [Russian original: Zh. Eksp. Teor. Fiz. 38, 1569 (1960)].

\bibitem{Peshkin} M. E. Peshkin and D. V. Schroeder, {\it An Introduction to Quantum Field Theory} (Addison-Willey Publishing Company, 1995).

\bibitem{Lee07} S. S. Lee, Phys. Rev. B {\bf76}, 075103(2007).

\bibitem{Grover14} T. Grover, D. N. Sheng and A. Vishwanath, Science {\bf344}, 280 (2014).

\bibitem{Yue10}	 Y. Yu and K. Yang, Phys. Rev. Lett. {\bf 105}. 150605(2010).

\bibitem{Zerf16} N. Zerf, C. H. Lin, and J. Maciejko, Phys. Rev. B 94, 205106 (2016).

\bibitem{Li17} Z.-X. Li, Y.-F. Jiang, and H. Yao, Phys. Rev. Lett. 119, 107202 (2017).

\bibitem{Jian17} S.-K. Jian, C.-H. Lin, J. Maciejko, and H. Yao, Phys. Rev. Lett. {\bf 118}, 166802 (2017).

\bibitem{Nagato09} Y. Nagato, S. Higashitani, and K. Nagai, J Phys. Soc. Jpn. {\bf 78} 123603 (2009).

\bibitem{Chung09} S. B. Chung and S.-C. Zhang, Phys. Rev. Lett. {\bf 103}, 235301 (2009).

\bibitem{Young12} S. M. Young, S. Zaheer, J. C. Y. Teo, C. L. Kane, E. J. Mele, and A. M. Rappe, Phys. Rev. Lett. {\bf108}, 140405 (2012).

\bibitem{SCbook} C Poole, H. Farach, R. Creswick, and R. Prozorov, Superconductivity, 3rd Edition (Elsevier, 2014).

\bibitem{Zhang09} H. Zhang, C.-X. Liu, X.-L. Qi, X. Dai, Z. Fang, and S.-C. Zhang, Nat. Phys. {\bf 5}, 438 (2009).

\bibitem{Gross74} D. J. Gross and A. Neveu, Phys. Rev. D {\bf10}, 3235 (1974).

\end{thebibliography}
\end{document}